\documentclass[a4paper,11pt]{article}
\pdfoutput=1 

\usepackage{jheppub} 

\usepackage[T1]{fontenc}
\usepackage[utf8]{inputenc}
\usepackage{braket}
\usepackage{subfig}
\usepackage{colortbl}
 \newcommand{\rowc}{\rowcolor[gray]{0.925}}
\usepackage{booktabs}

\title{\boldmath Vacuum Stability of Asymptotically Safe Two Higgs Doublet Models}
 \preprint{DO-TH 18/05}

 \author{Peter Schuh}
 \affiliation{Fakult\"at Physik, TU Dortmund, Otto-Hahn-Str. 4, D-44221 Dortmund, Germany}

\emailAdd{peter.schuh@udo.edu}

\abstract{We study different types of Two Higgs Doublet Models (2HDMs) under the assumption that all quartic couplings' beta functions vanish simultaneously at the Planck scale. The Standard Model seems to display this property almost accidentally, because the Higgs boson mass is close to 125 GeV. This also ties closely into the question of whether the theory is stable or metastable. We investigate if such ``fixed points'' can exist in various $\mathbb{Z}_2$-symmetric 2HDM subclasses, and if the theories that meet these conditions are phenomenologically viable, as well as vacuum stable. We find that the fixed point condition drastically reduces the parameter space of 2HDM theories, but can be met. Fixed points can only exist in type II and type Y models, in regions of large tan$\beta$, and they are only compatible with all existing experimental bounds if the $\mathbb{Z}_2$-symmetry is at least softly broken, with a soft breaking parameter of at least $M_{12}$ $>$ 70 GeV (380 GeV) for type Y (type II) models. The allowed region falls into the alignment limit, with the mixing angle combination $|\alpha - \beta| \approx\frac{\pi}{2}$. While there are both vacuum-stable and vacuum-unstable solutions, only the vacuum-unstable ones really agree with Standard-Model-like CP-even Higgs boson mass values of 125 GeV. The vacuum-stable solutions favour slightly higher values. While scenarios of asymptotically safe 2HDM exist, they cannot improve over the Standard Model regarding the question of vacuum stability.}

\begin{document} 
\maketitle
\flushbottom

\section{Introduction}

The discovery and the mass measurement of a Standard Model (SM)-like Higgs boson by ATLAS and CMS in 2012 \cite{Aad:2012tfa, Chatrchyan:2012xdj} so far rank among the most impactful events in this century's particle physics. It is an interesting situation that the Higgs mass of $m_H$ $=$ $(125.5 \pm 0.5)$ GeV lies right at the edge of the so-called stability bound \cite{EliasMiro:2011aa, Branchina:2013jra, Buttazzo:2013uya}. Extrapolating from the Higgs mass value to very short distances shows that the LHC result seems to hint at a quartic coupling of $\lambda$ $=$ 0 at Planck scale-like energies, and also the renormalisation group (RG) beta function $\beta_\lambda ( m_{Pl})$ $\sim$ 0.  

The argument also works in reverse: Before the LHC experiments had discovered a Higgs boson, calculations were performed to show that initial conditions of $\lambda$ $=$ 0 and $\beta_\lambda$ $=$ 0 at high scales naturally point to Higgs mass values around 125 GeV \cite{Shaposhnikov:2009pv}, as do the combination of a vanishing beta function at high scales and the experimental measurement of the top quark mass, or of a vanishing quartic coupling at high scales and the top quark mass \cite{Holthausen:2011aa}. The idea of vanishing beta functions suggests a link to the field of \textit{Asymptotic Safety} \cite{Weinberg:1980gg, Wetterich:1992yh, Percacci:2011fr}, in which RG flow fixed points play a critical role. Originally a concept for quantum gravity, Asymptotic Safety has in recent years become a point of interest in SM extensions, as a tool for UV completion or generalised renormalisability \cite{Bond:2016dvk, Bond:2017wut, Pelaggi:2017abg, Bond:2017lnq, Bond:2018oco, Barducci:2018ysr, Mann:2017wzh}. \\ 
A relation between vanishing quartic couplings and vanishing beta functions at high scales and the measured Higgs mass at the LHC may be coincidental. On the other hand, the question of vacuum stability remains. Experimental results suggest that the SM vacuum is metastable, although agreement on how strong a statement can be made has not yet been reached \cite{Bednyakov:2015sca}. Here, we study how models with an enlarged scalar sector behave in this regard. To this end, we look at Two Higgs Doublet Models (2HDMs) and investigate if the same properties can be found, and what ramifications the existence of fixed points can have for these models. Specifically, we examine 2HDMs that exhibit simultaneously vanishing quartic coupling beta functions $\beta_{\lambda_i}(\mu)$ at the Planck scale $m_{Pl}$ $=$ $1.2\cdot 10^{19}$ GeV. 

This paper is organized as follows: In Section \ref{2hdm}, general properties of 2HDMs are reviewed. A detailed outline of how the analyses are performed are then given in Section \ref{approach}. Sections \ref{z2} and \ref{m12} subsequently treat different types of 2HDMs, including the complete softly-broken $\mathbb{Z}_2$-symmetric model. An Appendix contains the complete two-loop beta functions of all 2HDM couplings used in this work.

\section{2HDM} \label{2hdm}
In this section, we briefly review the features of the general 2HDM, before reviewing the current state of bounds on the model from different sources.
\subsection{General properties of the 2HDM}
A 2HDM contains two SU(2) doublets $\Phi_1$, $\Phi_2$ \cite{Lee:1973iz}. The most general scalar potential takes the form:
{\small
\begin{eqnarray}
 V &=&  m_{11}^2 \Phi_1^\dagger \Phi_1 +m_{22}^2 \Phi_2^\dagger \Phi_2 + \left(M_{12}^2\Phi_1^\dagger\Phi_2 + h.c.\right) \nonumber \\
  &+&  \frac{1}{2}\lambda_1\left(\Phi_1^\dagger\Phi_1\right)^2 + \frac{1}{2}\lambda_2\left(\Phi_2^\dagger\Phi_2\right)^2 + \lambda_3\left(\Phi_1^\dagger\Phi_1\right)\left(\Phi_2^\dagger\Phi_2\right) + \lambda_4\left(\Phi_1^\dagger\Phi_2\right)\left(\Phi_2^\dagger\Phi_1\right) \nonumber \\
  &+&  \left[\frac{1}{2}\lambda_5\left(\Phi_1^\dagger\Phi_2\right)^2 
 +  \lambda_6\left(\Phi_1^\dagger\Phi_1\right)\left(\Phi_1^\dagger\Phi_2\right) +  \lambda_7\left(\Phi_2^\dagger\Phi_2\right)\left(\Phi_2^\dagger\Phi_1\right) + h.c.\right].
\end{eqnarray}}
In this notation, following \cite{Branco2012}, $m_{11}$, $m_{22}$, and $\lambda_1$ to $\lambda_4$ are real-valued, whereas $M_{12}$, $\lambda_5$, $\lambda_6$, and $\lambda_7$ are complex parameters. Of these 14 degrees of freedom, only eleven are physical. The rest can be absorbed by making use of the freedom of choice of bases for the SU(2) doublets $\Phi_i$.

For spontaneous symmetry breaking (SSB), both fields $\Phi_1$ and $\Phi_2$ are assigned a vacuum expectation value (VEV): $\braket{\Phi_i}_0$ $=$ $\begin{pmatrix} 0, & \frac{v_i}{\sqrt{2}} \end{pmatrix}^\top$, with $v_i$ related to the SM VEV $v\simeq 246$ GeV via 
\begin{align*}
 v_1^2 + v_2^2=v^2.
\end{align*}
The SU(2) doublets contain eight physical fields $\Phi_i$ $=$ $\begin{pmatrix} \phi_i^+, & \frac{(v_i + \rho_i + i \eta_i)}{\sqrt{2}} \end{pmatrix}^\top$, three of which are absorbed during SSB. The remaining physical Higgs bosons after rotating into mass eigenstates are the charged Higgs $H^\pm$, a pseudoscalar Higgs $A$ and two CP-even scalar Higgs $h$, $H$. The rotation angle diagonalizing the CP-even scalar mass matrix is conventionally called $\alpha$, the angle diagonalizing the charged and CP-odd bosons is called $\beta$. The latter angle $\beta$ also appears in the ratio of $\frac{v_2}{v_1}$ $\equiv$ tan$\beta$. 

In general, 2HDMs permit tree-level FCNCs. According to the Paschos-Glashow-Weinberg theorem \cite{Paschos:1976ay, Glashow:1976nt}, a necessary and sufficient condition for their absence is to have all fermions of the same charge and helicity couple to the same Higgs doublet. There are effectively only four different ways of distributing fermions to doublets, as $\Phi_1$ and $\Phi_2$ are inherently interchangable: A model in which all fermions couple to the same Higgs doublet (usually $\Phi_2$) is called \textit{type I}, a model where up-type quarks couple to $\Phi_2$ and down-type quarks couple to $\Phi_1$ is called \textit{type II}. Aligning the leptons with up-type instead of down-type quarks results in the so-called \textit{lepton-specific} and \textit{flipped} models, or \textit{type X} and \textit{type Y}, respectively. In practice, the different types are usually enforced through discrete $\mathbb{Z}_2$-symmetries. The exact charge assignments of Higgs and fermion fields are listed in Tab. \ref{typetab1} \cite{Aoki:2009ha}. For our purpose the leptons only contribute minor corrections when compared to the quarks, so the primary computational focus will be on type I and type II models.

\begin{table}
\begin{center}
\begin{tabular}{c|cccccc}
\toprule
\centering
 &  $\Phi_1$ & $\Phi_2$ & $u^R$ & $d^R$ & $l^R$ & $Q^l,L^l$ \\
\midrule

type I & -1 & 1 & 1 & 1 & 1 & 1  \\ \rowc
type II & -1 & 1 & 1 & -1 & -1 & 1 \\
type X & -1 & 1 & 1 & 1 & -1 & 1 \\ \rowc
type Y & -1 & 1 & 1 & -1 & 1 & 1 \\
\bottomrule
\end{tabular}
\caption{$\mathrm{Z}_2$ charge assignments of Higgs doublets and fermion fields for different 2HDM types.}
\label{typetab1}
\end{center}
\end{table}
The Yukawa Lagrangian for type I and type II 2HDMs are hence given by:
\begin{align}
-\mathcal{L}_Y^I &= \left(\bar{Q^l_i}\Phi^2Y^d_{ij}d^R_j + \bar{Q^l_i}\tilde{\Phi}^2Y^u_{ij}u^R_j + \bar{L^l_i}\Phi^2Y^l_{ij}l^R_j\right) + h.c.,\\
 -\mathcal{L}_Y^{II} &= \left(\bar{Q^l_i}\Phi^2Y^d_{ij}d^R_j + \bar{Q^l_i}\tilde{\Phi}^1Y^u_{ij}u^R_j + \bar{L^l_i}\Phi^1Y^l_{ij}l^R_j\right) + h.c.,
\end{align}
where $Y^{u,d,l}$ are the Yukawa matrices for up-, down-, and lepton type particles, $Q^l$, $L^l$, $u^R$, $d^R$, and $l^R$ are left- and right-handed quark and lepton fields respectively, and $i,j$ denote the generations in flavour space. In our calculations, only the dominant $y_{33}$ entries generated by the top quark, the bottom quark, and the tau lepton respectively, will be considered. Thus, the Yukawa matrices are assumed to have the simplified structures $Y^u$ $=$ $ \textrm{diag}(0,0,\lambda_t)$, $Y^d$ $=$ $\textrm{diag}(0,0,\lambda_b)$ and $Y^l$ $=$ $\textrm{diag}(0,0,\lambda_\tau)$.

Under the $\Phi_1 \rightarrow -\Phi_1$  $\mathbb{Z}_2$ symmetry mentioned above, it follows that $\lambda_6$ $=$ $\lambda_7$ $=$ 0, which leads to a mass matrix for the CP-even neutral scalars of the form: 
\begin{align}
M_{h/H}^2 &= 
\begin{pmatrix}
 m_{11}^2  + \frac{3}{2}\lambda_1v_1^2 + \frac{3}{2}\lambda_{345}v_2^2 & - \textrm{Re}(M_{12}^2) + \lambda_{345}v_1v_2 \\  -\textrm{Re}(M_{12}^2) + \lambda_{345}v_1v_2 & m_{22}^2 + \frac{3}{2}\lambda_2v_2^2  + \frac{3}{2}\lambda_{345}v_2^2
 \end{pmatrix},
 \label{cpevenmatrix}
\end{align} 
with $\lambda_{345}$ $=$ $\lambda_3 + \lambda_4 + \textrm{Re}(\lambda_5)$. The terms $m_{11}$ and $m_{22}$ can be eliminated using the minimum conditions from SSB, that is $\frac{\partial V}{\partial v_i}$ $=$ 0:
\begin{align}
 m_{11}^2v_1 - \textrm{Re}(M_{12}^2)v_2 + \frac{\lambda_1}{2}v_1^3 + \frac{\lambda_{345}}{2}v_1v_2^2 &= 0, \label{mineq1}\\
 m_{22}^2v_2 - \textrm{Re}(M_{12}^2)v_1 + \frac{\lambda_2}{2}v_2^3 + \frac{\lambda_{345}}{2}v_1^2v_2 &= 0.
 \label{mineq2}
\end{align}
The charged and the pseudoscalar Higgs mass matrices are given by:

\begin{align}
  M_{H^\pm}^2 &=  \frac{v^2}{v_1v_2}\Big(\textrm{Re}(M_{12}^2) - \frac{\lambda_4 + \textrm{Re}(\lambda_5)}{2}v_1 v_2\Big)
 \begin{pmatrix}
\frac{v_2}{v_1} & -1 \\ -1 & \frac{v_1}{v_2}
 \end{pmatrix},\\
 M_{A}^2 &= \Big(\textrm{Re}(M_{12}^2) -  \textrm{Re}(\lambda_5)v_1 v_2\Big)
 \begin{pmatrix}
\frac{v_2}{v_1} & -1 \\ -1 & \frac{v_1}{v_2}
 \end{pmatrix}.
 \label{masses2}
\end{align}
Both have one zero eigenvalue, corresponding to the charged and the pseudoscalar Goldstone boson, respectively. The pseudoscalar mass vanishes for $M_{12}$ $=$ $\lambda_5$ $=$ 0, because of an additional accidental spontaneously broken U(1)-symmetry.

In 2HDMs, to be vacuum-stable the potential needs to be bounded from below in all directions. This is the case if and only if the following set of inequalities is met \cite{Deshpande:1977rw}:
\begin{align}
 \lambda_1 &> 0, & \lambda_2 &> 0, \nonumber \\
 \lambda_3 + \sqrt{\lambda_1\lambda_2} & > 0, & \lambda_3 + \lambda_4 + \sqrt{\lambda_1\lambda_2} &> |\lambda_5|.
 \label{stabilitycond}
\end{align}
Unlike the SM, theories with more than one Higgs doublet can display a range of different vacuum configurations \cite{Branchina:2018qlf}: Not only can there be more than one minimum at the same time, but the minima can also be of CP breaking type, when the VEVs have a relative complex phase, or of charge breaking type, with one VEV carrying an electric charge. It has however been shown that minima of different types (i.e. CP-breaking, charge-breaking, or normal) cannot exist simultaneously within the same model \cite{Ferreira:2004yd, Barroso:2005sm, Barroso:2007rr}. By requiring the model to fulfil the minimum conditions for normal-type minima given by Eqs. \eqref{mineq1} and \eqref{mineq2}, it is therefore assured that the absolute minimum of the theory is also normal. It only remains to be checked if the minimum at $v=246$ GeV is global, or if there is another, deeper one.

\subsection{Limits on 2HDM Parameter Space} \label{limits}
While the 2HDM is a relatively simple SM extension, it still contains up to eleven new free parameters (six in the type II models studied below). On the other hand, the model's high popularity means that its parameter space has been comprehensively explored and constrained from both the theoretical and the experimental side, and in particular by recent LHC data \cite{Barroso:2013zxa, Coleppa:2013dya, Sirunyan:2018koj}. At this point, we briefly review current bounds, more thorough discussions of different aspects can be found for example in \cite{Eberhardt:2017ulj, Chowdhury:2017aav, Dorsch:2016tab, Cheon:2012rh, Eberhardt:2013uba, Chowdhury:2015yja, Chakrabarty:2014aya}.

In essence, constraints on the 2HDM parameter space can be sorted into three categories: Theory bounds are generated by requiring the model to possess certain features, commonly referred to as \textit{positivity} (the Higgs potential must be bounded from below, cf. Eq. \eqref{stabilitycond}), \textit{perturbativity} (quartic couplings must not be large), and \textit{unitarity} (of the S-matrix of 2$\rightarrow$2 scattering amplitudes) \cite{Lee:1977yc, Grinstein:2015rtl}.
Secondly, there are mass bounds on the physical Higgs bosons from signal strength data by the ATLAS and CMS collaborations. These searches have confirmed the existence of a 125 GeV CP-even scalar eigenstate, and they also show that this boson couples to vector bosons and fermions in a very SM-Higgs-like fashion \cite{ATLAS:2014yka, CMS:2014ega, ATLAS:2018uoi, CMS:2018lkl}. Furthermore, the absence of heavier resonances so far translates to mass bounds for the other Higgs eigenstates. 
Lastly, there are implications for the 2HDM from flavour physics \cite{Enomoto:2015wbn}. Most notably, $\mathcal{B}(b\rightarrow s \gamma)$ measurements exclude charged Higgs masses smaller than $m_{H^+}$ $=$ 580 GeV \cite{Misiak:2017bgg} in type II/type Y models, lower bounds on tan$\beta$ can be extracted from $B_s$ mass differences and leptonic decays \cite{Haller:2018nnx}.

Together, these bounds can be combined to make a number of statements: 
The masses of the three additional Higgs bosons all must be large, the mass differences between them, however, small. The rotation angles must fulfil $|\beta - \alpha|$ $\approx$ $\frac{\pi}{2}$, ensuring that the mass basis of the CP-even scalar states aligns with the SM gauge eigenbasis. These features are thus usually referred to as \textit{alignment limit} \cite{Gunion:2002zf, Das:2015mwa, Carena:2013ooa, Dev:2014yca}.
It should be noted that some studies have used fine-tuning arguments to impose stronger bounds on tan$\beta$, and successively to the heavy Higgs boson masses \cite{Chowdhury:2017aav}. Since the RG methods employed in this work contain a certain degree of tuning by design, they offer an alternative as to why these large tan$\beta$ regions may yet be phenomenologically viable. As a consequence, our mass bounds are slightly more conservative than some.

\section{Solving the Fixed Point Equations}\label{approach}

We pursue the question whether 2HDMs support ''fixed points`` at the Planck scale in the same way the SM does, and if the resulting models are vacuum-stable. 
The fixed point condition reads:
\begin{align}
 \beta_{\lambda_i}(m_{Pl}) &= 0 \quad \forall i,
 \label{fpe}
\end{align}
i.e., the beta functions of all quartic couplings $\lambda_i$ present in the scalar potential are to have a root at the Planck mass $m_{Pl}$. Because of contributions from gauge and Yukawa couplings, the condition of $\beta_{\lambda_i}$ $=$ 0 is not technically sufficient to define a fixed point, nor does it necessarily lead to an asymptotically safe theory by itself; for recent progress in BSM model building see e.g.  \cite{Bond:2017wut}. Still, because of similarities to the SM case and for convenience, the terms \textit{fixed point} and \textit{fixed point condition} will be used in this context, effectively interpreting effects disturbing the equilibrium into the realm of beyond the Planck scale physics.

While in the SM there is only one quartic Higgs coupling $\lambda$, the 2HDM potential with a $\mathbb{Z}_2$ symmetry protecting flavour conservation can contain up to five quartic terms (one of which may be complex). This means that compared to the SM, the fixed point condition has a much higher impact in terms of limiting the parameter space of the theory.

The search for fixed points comes down to solving the system of differential equations given by the beta functions of the running couplings of the theory. It involves the gauge couplings $g_1$, $g_2$, $g_3$, the quartic scalar couplings $\lambda_i$ and the Yukawa couplings $\lambda_t$, $\lambda_b$, $\lambda_\tau$. The complete two-loop expressions for the most general beta functions used are calculated with the Mathematica package SARAH \cite{Staub:2008uz, Staub:2013tta}, and listed in Appendix A. Since the coefficients $m_{11}$ and $m_{22}$ of the dimension-two-operators do not appear directly in the beta functions of any other couplings, $m_{ii}$ can be ignored at this point and determined with help of the minimum conditions at the electroweak scale, see Eqs. (\ref{mineq1}) and (\ref{mineq2}). The soft breaking parameter $M_{12}$ also does not appear in the beta functions of quartic, gauge, or Yukawa couplings, and will be treated as a free parameter.

While the quartic couplings are already fixed implicitly by \eqref{fpe}, the remaining initial conditions are given explicitly at low scales: Both gauge couplings and Yukawa couplings can be determined from experimental measurements. The $\overline{\textrm{MS}}$ gauge coupling initial conditions for $g_1$ and $g_2$ are calculated using the fine structure constant $\alpha^{-1}(M_Z)=127.95 \pm 0.017$ and the weak mixing angle $sin^2\theta_W=0.23129 \pm 5\cdot 10^{-5}$ \cite{Mohr:2015ccw,Agashe:2014kda} to:
\begin{align*}
  g_1(M_Z) &= 0.35, & g_2(M_Z) &=0.65, & g_3(M_Z) &=1.2.
\end{align*}
Uncertainties on gauge coupling initial values are small enough to be inconsequential.
The relations between Yukawa couplings and quark masses are model-dependent. In a type II model, the Yukawa couplings are related to the quark masses by the tree-level relations:
\begin{align*}
 \lambda_t(m_t) &=\frac{\sqrt{2}m_t}{v_2}, &  \lambda_b(m_b) &=\frac{\sqrt{2}m_b}{v_1}, & \lambda_\tau(m_\tau) &=\frac{\sqrt{2}m_\tau}{v_1},
  \end{align*}
  while in the type I model the bottom quark Yukawa coupling is instead determined by the other VEV: $\ \lambda_b(m_b) =\frac{\sqrt{2}m_b}{v_2}$.
  The $\overline{MS}$ quark masses used are $m_b(m_b)$ $=$ 4.18 $\pm$ 0.03 GeV and $m_t(m_t)$ $=$ 160$^{+4.8}_{-4.3}$ GeV, the $\tau$-lepton mass is given by $m_\tau(m_\tau)$ $=$ 1.78 GeV \cite{Agashe:2014kda}. For the purpose of this paper, it is assumed that all non-SM effects only affect the running from the electroweak scale onwards. In other words, the Yukawa couplings are run up to $M_Z$ under SM-like conditions, at which point the tan$\beta$-enhancement is switched on. The effective initial values used are thus:
\begin{align*}
 \lambda_t(M_Z) &=\frac{0.95}{\text{sin}\beta}, &  \lambda_b(M_Z) &=\frac{0.176}{\text{cos}\beta}, & \lambda_\tau(M_Z) &=\frac{0.98}{\text{cos}\beta}. \label{enhance}
\end{align*}
The so defined initial value problem is solved numerically.

With the full RG flow of all couplings known, their low scale values are used to determine the mass spectrum of the theory using the matrices given in Section \ref{2hdm}. The results depend on the parameters treated as free (in these cases tan$\beta$ and later $M_{12}$) and on the experimentally determined coupling initial conditions, but beyond this are a direct consequence of the theory itself and the fixed point assumption.

With all couplings known at all scales, the question of vacuum stability can also be answered: For a solution to be vacuum-stable, the quartic couplings must fulfil the inequalities given by Eq. (\ref{stabilitycond}) at all energy scales up to $\mu$ $=$ $m_{Pl}$.

\section{The CP-conserving 2HDM with $\mathbb{Z}_2$ Symmetry} \label{z2}
We study a 2HDM with a discrete $\mathbb Z_2$ symmetry ($\Phi_1 \rightarrow -\Phi_1$, $\Phi_2 \rightarrow \Phi_2$) introduced in order to ensure CP-conservation in the scalar sector. The scalar potential reads:

\begin{eqnarray}
 V & = &  m_{11}^2 \Phi_1^\dagger \Phi_1 +m_{22}^2 \Phi_2^\dagger \Phi_2 +\frac{1}{2}\lambda_1\left(\Phi_1^\dagger\Phi_1\right)^2  
 +   \frac{1}{2}\lambda_2\left(\Phi_2^\dagger\Phi_2\right)^2 \nonumber \\ & + &  \lambda_3\left(\Phi_1^\dagger\Phi_1\right)\left(\Phi_2^\dagger\Phi_2\right) 
 +  \lambda_4\left(\Phi_1^\dagger\Phi_2\right)\left(\Phi_2^\dagger\Phi_1\right) 
 +  \frac{1}{2}\lambda_5\left[\left(\Phi_1^\dagger\Phi_2\right)^2 + \left(\Phi_2^\dagger\Phi_1\right)^2\right],
\end{eqnarray}
with real-valued mass parameters $m_{ii}$ and quartic couplings $\lambda_i$. The restriction Im($\lambda_5$) $=$ 0 does not follow immediately from the $\mathbb Z_2$ symmetry, but can be assumed in this case without loss of generality due to the structure of the beta functions:
$\lambda_5$ always appears in the other quartic couplings' beta functions in the form of the norm squared, $|\lambda_5|^2$, and the function $\beta_{\lambda_5}$ is proportional to $\lambda_5$, with the remainder being comprised by real-valued terms only. It follows that the function $\beta_{\lambda_5}$ must have a constant phase. It is also easy to see that in these circumstances for every function $\lambda_5(\mu)$ that solves the system of differential equations, a phase-shifted $e^{i\theta}\cdot \lambda_5(\mu)$ is also a solution for every constant phase $\theta$. Therefore, it suffices to look at the real values for $\lambda_5$ when searching for fixed points.

The different Yukawa Lagrangians in type I and type II models lead to differences in the beta functions, as the Yukawa terms will appear in the running of different couplings. As an example, in the type I model one-loop beta function of quartic coupling $\lambda_1$ can be written as:
\begin{align}
\beta_{\lambda_1}^{1l} &= \frac{1}{64\pi^2} \Big(6 \text{$g_1$}^2\text{$g_2$}^2 +  \Big(3\text{$g_2$}^2-6 \text{$\lambda_1$}\Big)^2 + 3\Big(\text{$g_1$}^2-2 \text{$\lambda_1$}\Big)^2\nonumber \\
  &+8 \Big(6 \text{$\lambda_1$}^2+ \text{$\lambda_3$}^2+ \Big(\text{$\lambda_3$} + \text{$\lambda_4$} \Big)^2+\text{$\lambda_5$}^2\Big)\Big).
\end{align}
The right hand side is expressed as a sum of perfect squares. Furthermore, it does not contain any Yukawa terms, as in type I models, $\Phi_1$ does not couple to any fermions. As was also discussed in \cite{Bond:2016dvk}, Yukawa couplings are often times indispensable for enabling fixed points, as they can be the only terms with a negative sign in the beta functions. The consequence of this is that the $\beta_{\lambda_1}^{1l}$ above will never vanish, and type I models are excluded as viable candidates, when looking for 2HDMs with fixed points. As a corollary, models like the one discussed in \cite{Falkowski:2015iwa} extending the SM by a singlet can also not support fixed points in their new quartic coupling.

\begin{figure}
 \centering
      \includegraphics[width=0.7\textwidth]{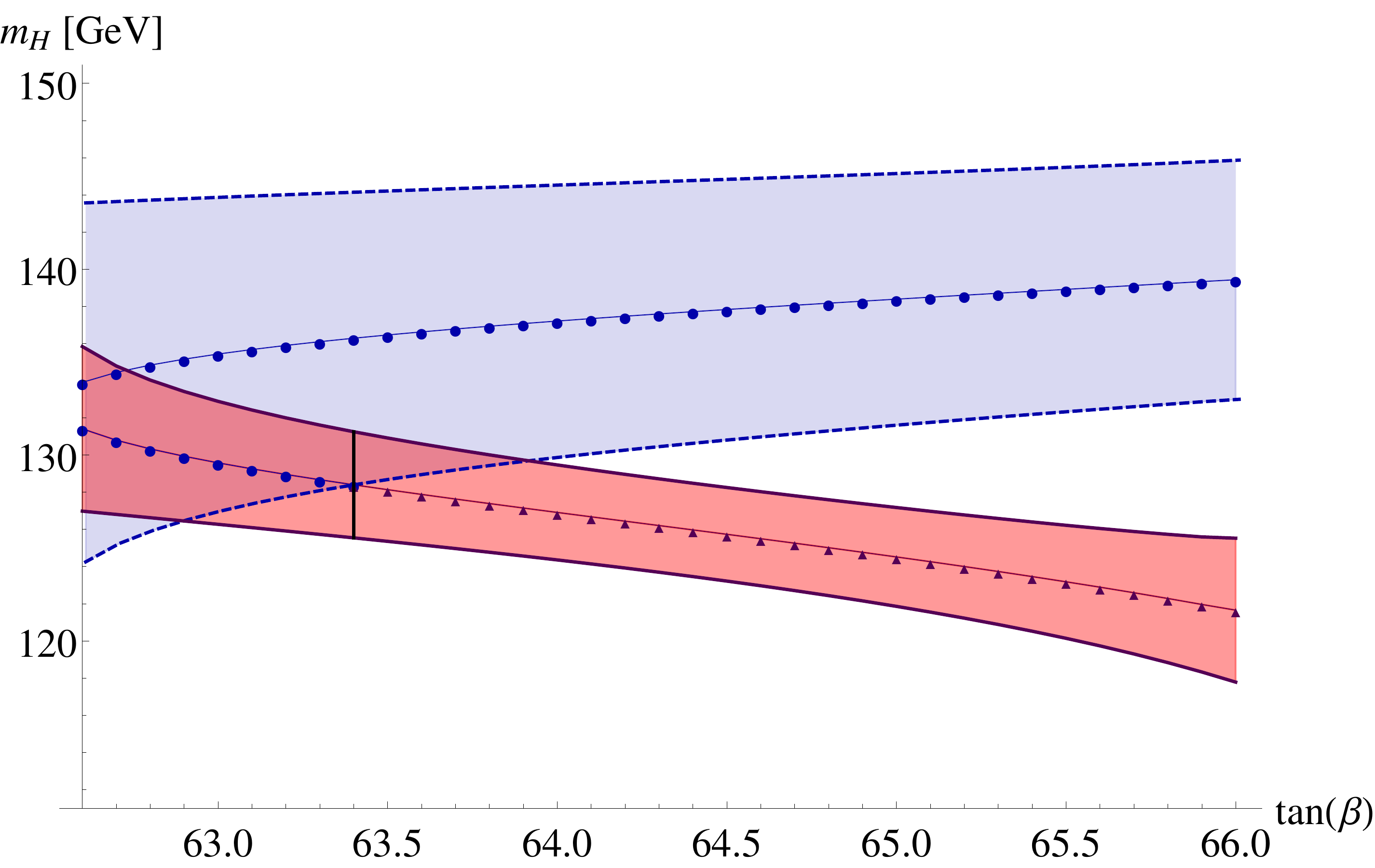}
  \captionof{figure}{Mass eigenstates of the heavier CP-even neutral scalar Higgs boson $H$ in the type II $\mathbb{Z}_2$-symmetric model. Vacuum-stable solutions are shown as blue circles with dashed borders, vacuum-unstable solutions as violet triangles, with the black vertical line marking the transition point. The uncertainty estimate is given by the difference between one-loop and two-loop results.}
\label{fig3}
 \end{figure}

\begin{figure}
 \centering 
      \includegraphics[width=0.7\textwidth]{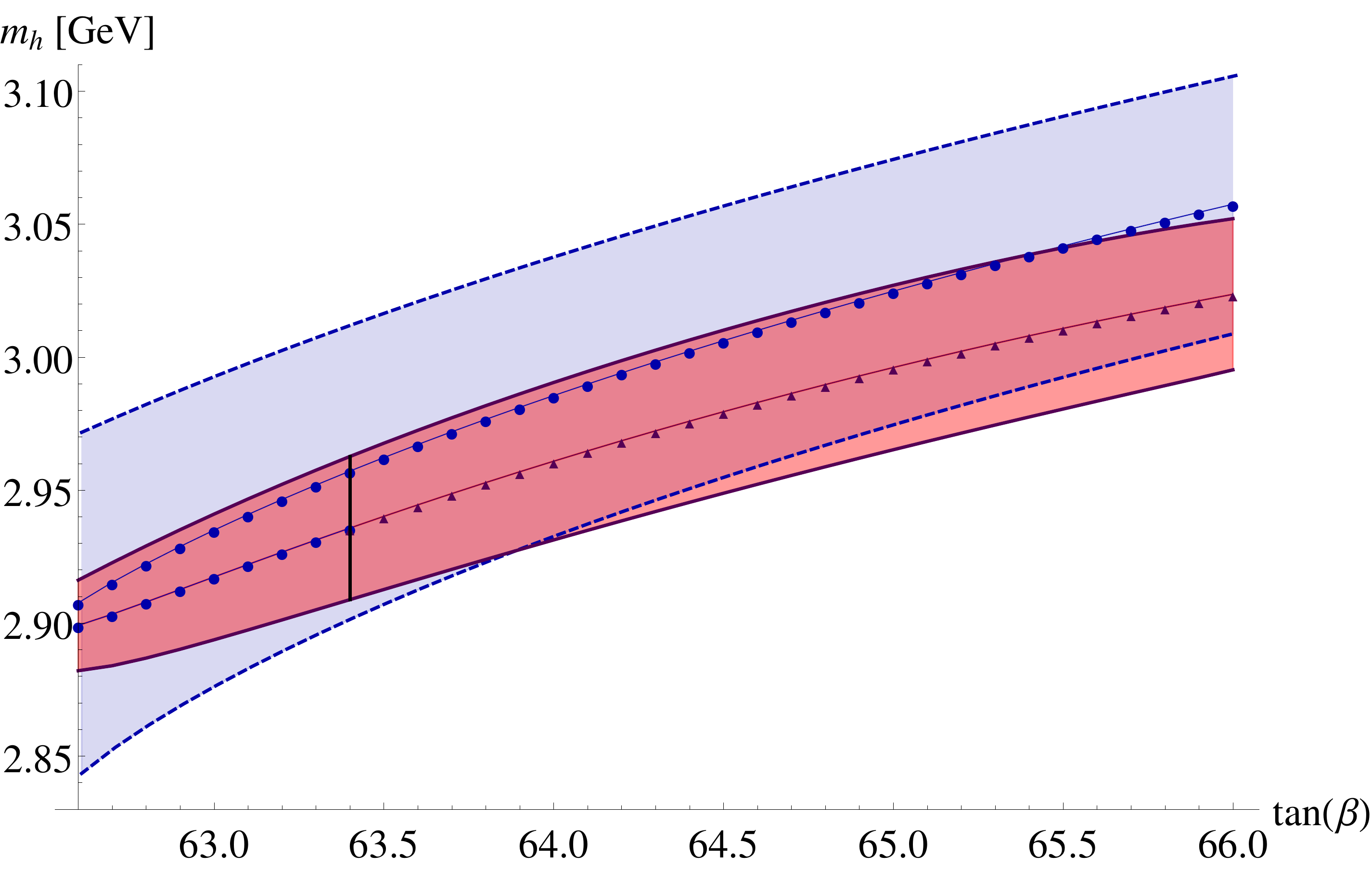}
  \captionof{figure}{Mass eigenstates of the light CP-even neutral scalar Higgs boson $h$ in the type II $\mathbb{Z}_2$-symmetric model, as labeled in Fig. \ref{fig3}.}
\label{fig3b}
 \end{figure}

\begin{figure}
 \centering 
    \includegraphics[width=0.7\textwidth]{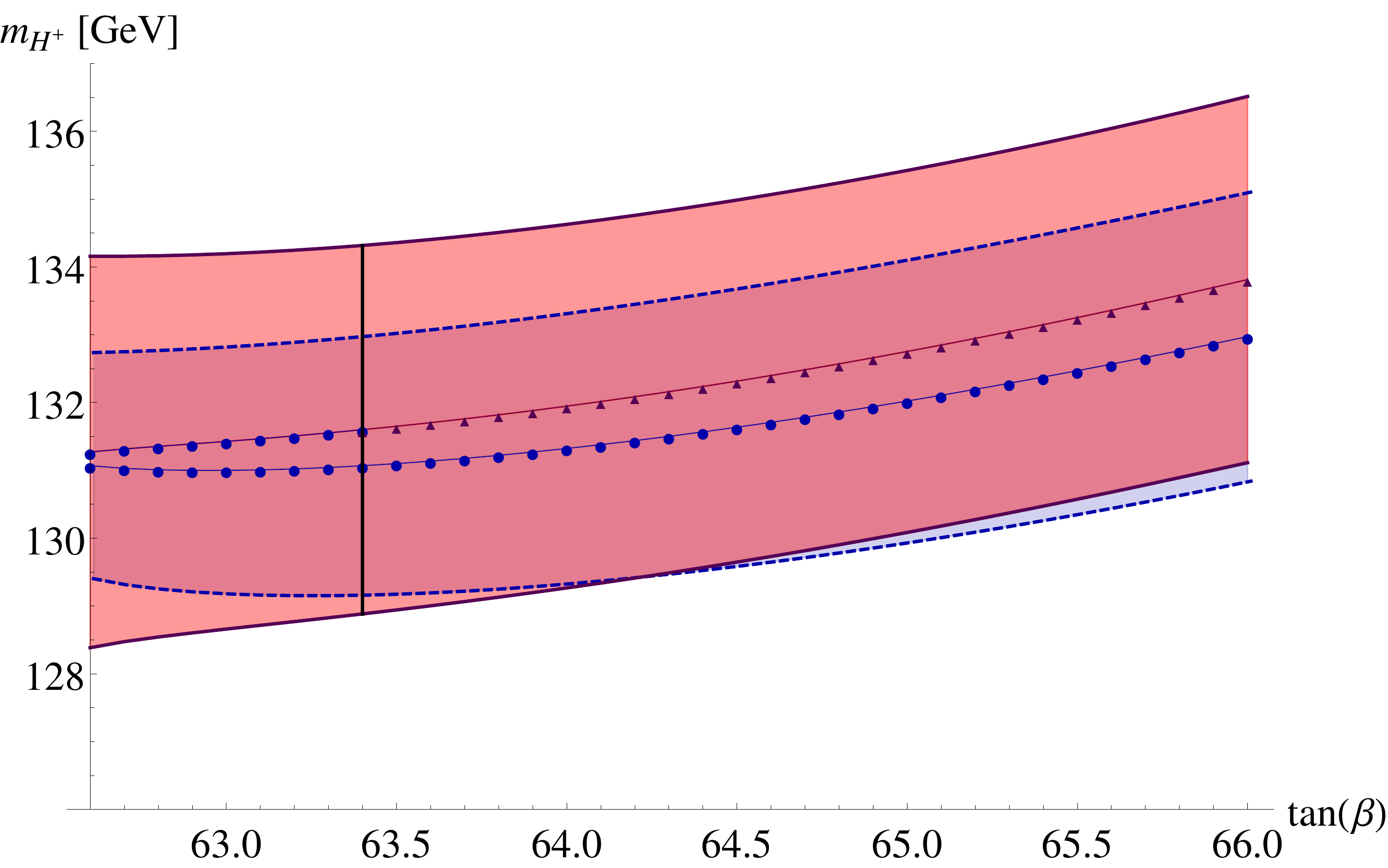}
\captionof{figure}{Mass eigenstates of the charged Higgs boson $H^+$ in the type II $\mathbb{Z}_2$-symmetric model, as labeled in Fig. \ref{fig3}.}
\label{fig3c}
 \end{figure}

The full two-loop beta functions for the type II $\mathbb{Z}_2$-symmetric 2HDM are given in Appendix A. In this model, tan$\beta$ is a free parameter that fixes the exact starting conditions. For fixed points to exist, Yukawa contributions to both Higgs field have to be big enough. This translates into a lower bound on tan$\beta$, as the bottom-type Yukawa couplings in a type II model are $(\sin\beta)^{-1}$-enhanced compared to the SM (cf. Eq. \eqref{enhance}). On the other hand, tan$\beta$ cannot be too large either, or the running Yukawa couplings will become divergent below the Planck scale. Accordingly, there is a tan$\beta$-interval in which fixed points can exist. In this two-loop framework, fixed points can be found for:
\begin{align}
 \textrm{tan$\beta$} \in \left[60,70\right].
 \label{interval}
\end{align}
The beta functions being of polynomial form, there is usually a finite set of fixed point solutions. We find that there are two branches: One branch is vacuum-stable for all values of tan$\beta$, the other one breaks the condition $\lambda_2 > 0$ for certain values of tan$\beta$, rendering the corresponding potential unbounded from below. In either case, the values of $\lambda_i$ can be translated into a spectrum of masses of the physical Higgs bosons. The masses for all three physical bosons in this model are shown in Figures \ref{fig3}, \ref{fig3b} and \ref{fig3c}, with the vacuum-stable cases shown in blue, and the vacuum-unstable solutions marked violet. The critical tan$\beta$ value for stability in the lower branch is marked by a red vertical line. An estimate for the theoretical uncertainty is given by the difference between one-loop and two-loop results.

The CP-even neutral scalar mass eigenstates are split apart wide due to the large parameter tan$\beta$: Without significant mixing, one of the eigenstates is of the order of $v_1$, the other of the order $v_2$. The heavier CP-even eigenstate takes on mass values of around 125-130 GeV for the (mostly) vacuum-unstable branch, and 135-140 GeV for the stable one. 
The lighter CP-even eigenstate lies in the O(1) GeV region, which is ruled out experimentally. This will be addressed in Section \ref{m12}.\\
The heavier CP-even eigenstate in the vacuum-unstable case has roughly the correct mass to be considered as a candidate for a SM-like Higgs.  However, while it is theoretically possible that the observed Higgs boson at 125 GeV is the \textit{heavier} of the two CP-even eigenstates, this configuration is heavily disfavoured by experimental observations, due to strong bounds from the $H \rightarrow hh$ decay \cite{Ferreira:2012my}. It is therefore usually assumed in 2HDMs that the 125 GeV Higgs is the lighter of the two CP-even neutral eigenstates.\\
The charged Higgs boson mass is below 200 GeV, which falls into the regions excluded by $\bar{B}\rightarrow X_s\gamma$ measurements, as mentioned in Section \ref{limits}. 
The pseudoscalar mass is not shown, because it vanishes completely: As it turns out, all fixed point solutions contain $\lambda_5$ $\equiv$ 0. In this case (with $M_{12}$ forbidden by the $\mathbb{Z}_2$-symmetry), this means that the model displays an accidental U(1)-symmetry which forces the pseudoscalar into the role of a pseudo-Goldstone boson, and hence to become massless.
 The mixing angle $\alpha$ is close to zero in all cases.

\begin{figure}
	\centering
	\includegraphics[width=0.7\textwidth]{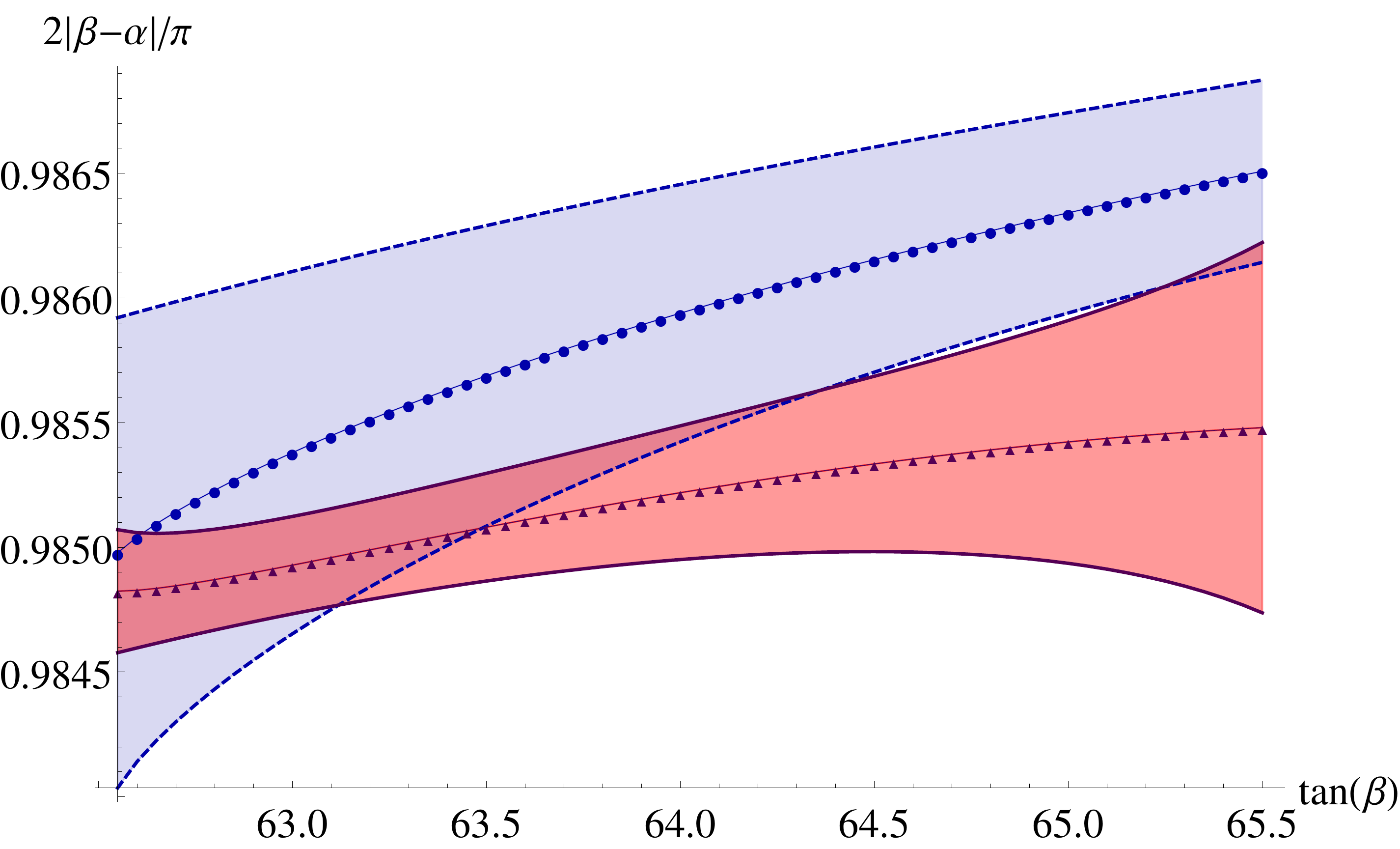}
	
	\caption{Mixing angle difference $|\beta-\alpha|$ with difference between one-loop and two-loop fixed point results as uncertainty. The vacuum-stable and -unstable branches are marked as labeled in Fig. \ref{fig3}}
	\label{figalphabeta}
\end{figure}

Together with the large tan$\beta$ values, this ensures that the type II alignment limit condition of $|\beta - \alpha| \sim \frac{\pi}{2}$ (cf. Section \ref{limits}) is always met, as is shown in Fig. \ref{figalphabeta}.

To summarise, the mass spectrum produced by the vacuum-stable solutions to the fixed point equations in the  $\mathbb{Z}_2$-symmetric 2HDM exhibits an SM-Higgs candidate in the vacuum-unstable branch, but is excluded by experimental observations because of the remaining boson spectrum. We fix this problem in the next section. Performing the same analysis in a type-Y 2HDM leads to generally analogous results at slightly elevated tan$\beta$-values.\\

\section{The 2HDM With Softly Broken $\mathbb{Z}_2$ } \label{m12}
To allow for larger, phenomenologically viable masses for $m_h$, $m_{H^+}$, and $m_A$, it is necessary to go beyond the $\mathbb{Z}_2$-symmetric model. The least invasive way to generate heavier masses is to include the so-called \textit{softly-broken} $\mathbb{Z}_2$-symmetric 2HDMs. The assumption of a $\mathbb{Z}_2$-symmetry under the transformation $\phi_2 \rightarrow -\phi_2$ is not completely dropped, but a mass term $M_{12}^2(\Phi_1^\dagger\Phi_2+\Phi_2^\dagger \Phi_1)$ mixing between the two Higgs field is allowed.

\begin{figure}
	\centering
	\includegraphics[width=0.7\textwidth]{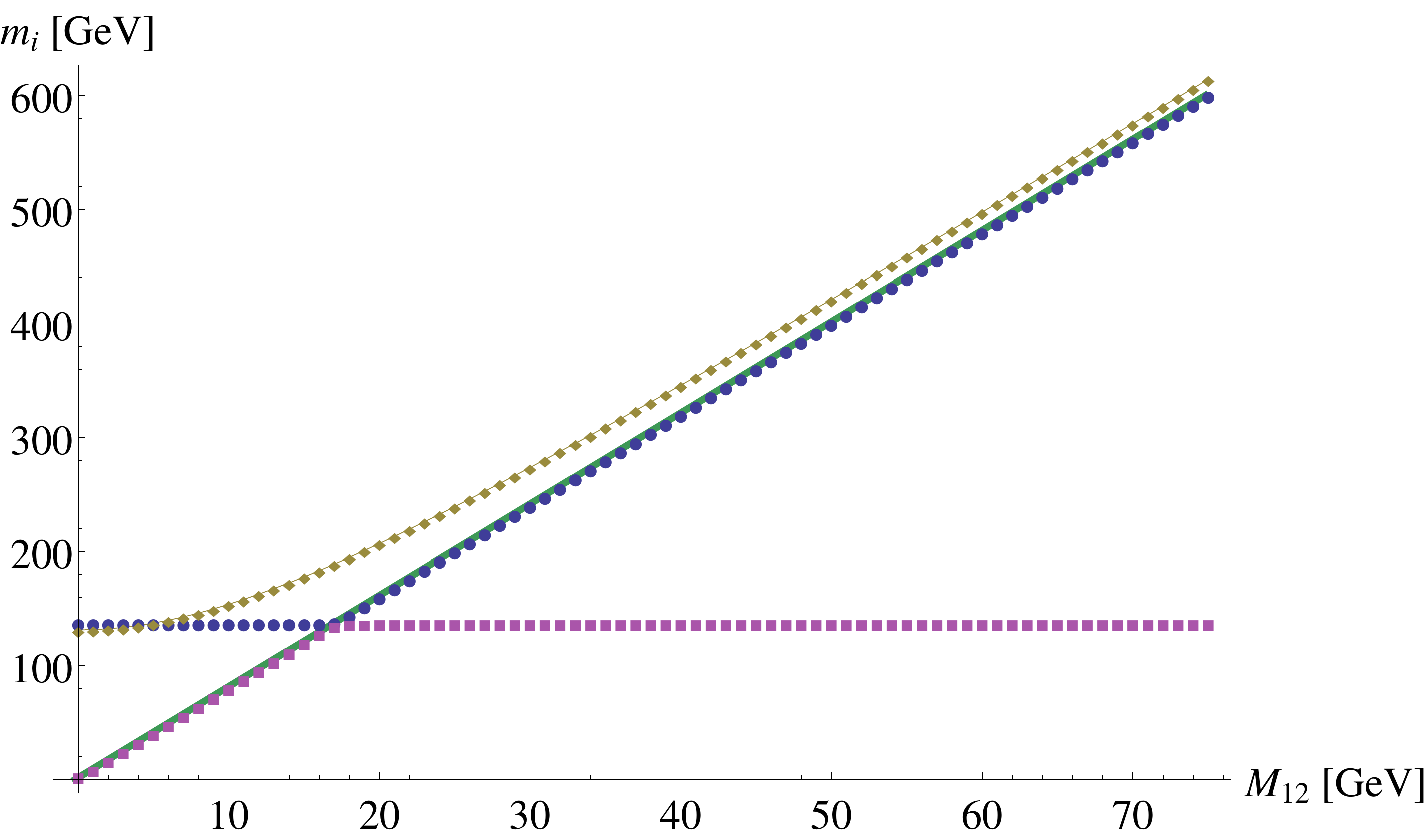}
	
	\caption{Higgs boson masses for a softly broken $\mathbb{Z}_2$-symmetry for tan$\beta$ $=$ 64 against $M_{12}$ in GeV for the vacuum-unstable branch of solutions. The masses shown correspond to the CP-even scalars $m_h$, $m_H$ (violet squares, blue circles), the charged Higgs $m_{H^+}$ (yellow diamonds) and the pseudoscalar $m_A$ (green line).} 
	\label{lc4}
\end{figure}

Like the other mass parameters, $M_{12}$ does not appear in any quartic, gauge, or Yukawa beta function. It can instead be treated as a free parameter. For this reason, $M_{12}$ does not influence the fixed point search itself. Phenomenologically, on the other hand, the mixing parameter can have a big impact, especially in the case of $\lambda_5$ $=$ $0$ observed in our models. The additional global U(1) symmetry is now broken by non-vanishing $M_{12}$-terms, which allows the pseudoscalar Higgs boson to acquire mass. Additionally, three of the other four bosons grow approximately linear with $M_{12}$, which allows them to evade experimental constraints. The influence of $M_{12}$ on the different boson masses is illustrated in Fig. \ref{lc4}.\\
As can be seen in the plot, the CP-even neutral scalar eigenvalues (blue/violet) depend on $M_{12}$ in different ways (cf. Eq.\eqref{cpevenmatrix}). The SM-like eigenstate (originally $m_H$) only shows a minor dependence, and hardly changes even for very large values of $M_{12}$. For the originally smaller eigenstate however, $M_{12}$ can easily become the dominating contributor. While the original mass of this state was mainly generated by a small VEV $v_2$, it soon starts to grow almost linearly with $M_{12}$, surpassing the mass of the former heavier eigenstate in a level crossing at values of roughly $M_{12}$ $\approx$ 20 GeV.

Both the charged Higgs (yellow) and the pseudoscalar Higgs (green) also grow together with $M_{12}$, and adopt an asymptotically linear behaviour as $M_{12}$ becomes large. For the pseudoscalar, the linear dependence is actually exact as long as $\lambda_5$ $=$ 0, as it is the case with our fixed point solutions. The mass eigenvalue in \eqref{masses2} then simplifies to:
\begin{align}
 m_A &= \frac{M_{12}}{\sqrt{\sin\beta\cos\beta}}\sim M_{12} \sqrt{\textrm{tan}\beta}.
\end{align}
For this reason the pseudoscalar mass $m_A$ in Fig. \ref{lc4} is shown as a straight line.

\begin{figure}
	\subfloat{%
		\includegraphics[width=0.49\textwidth]{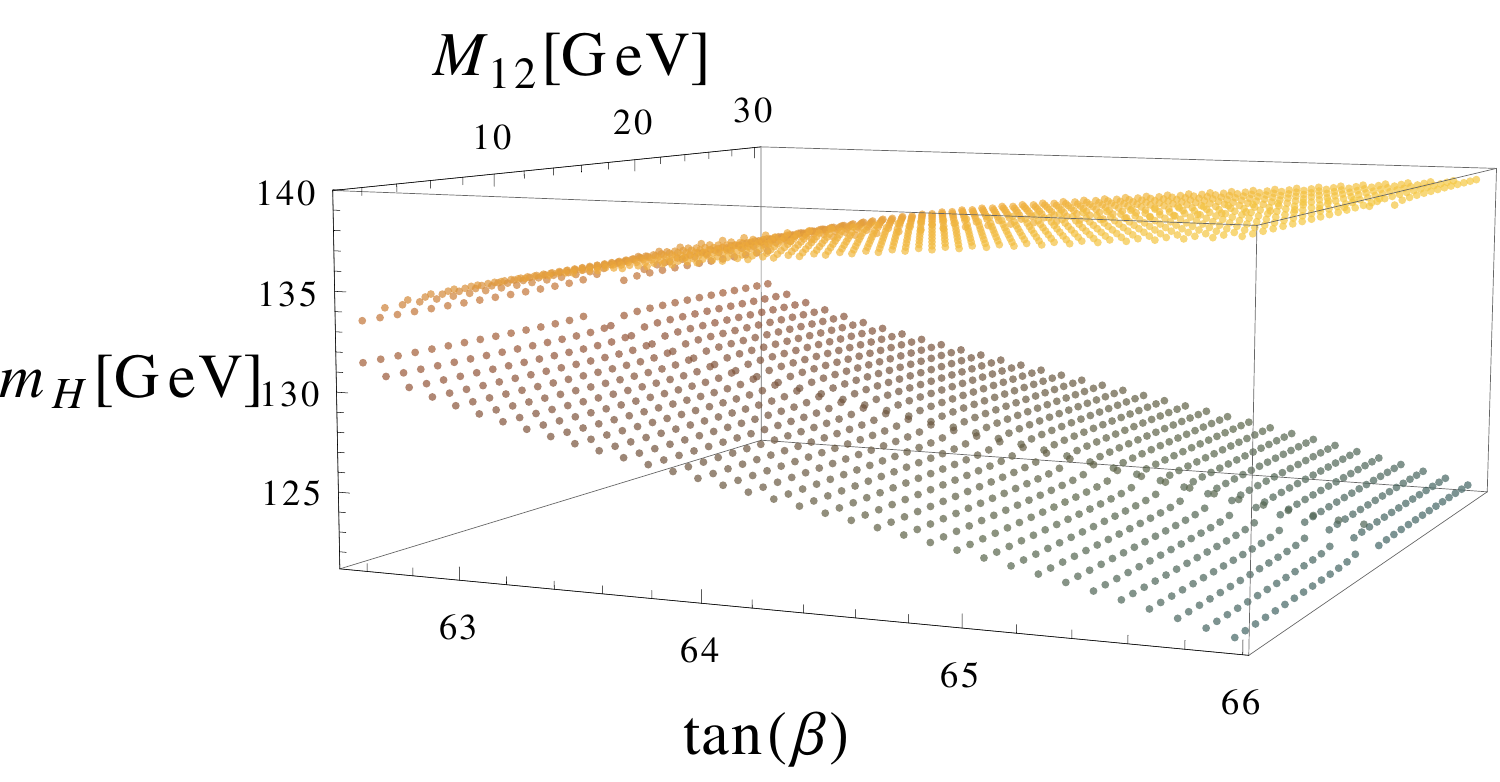}
	}
	\hfill
	\subfloat{%
		\includegraphics[width=0.5\textwidth]{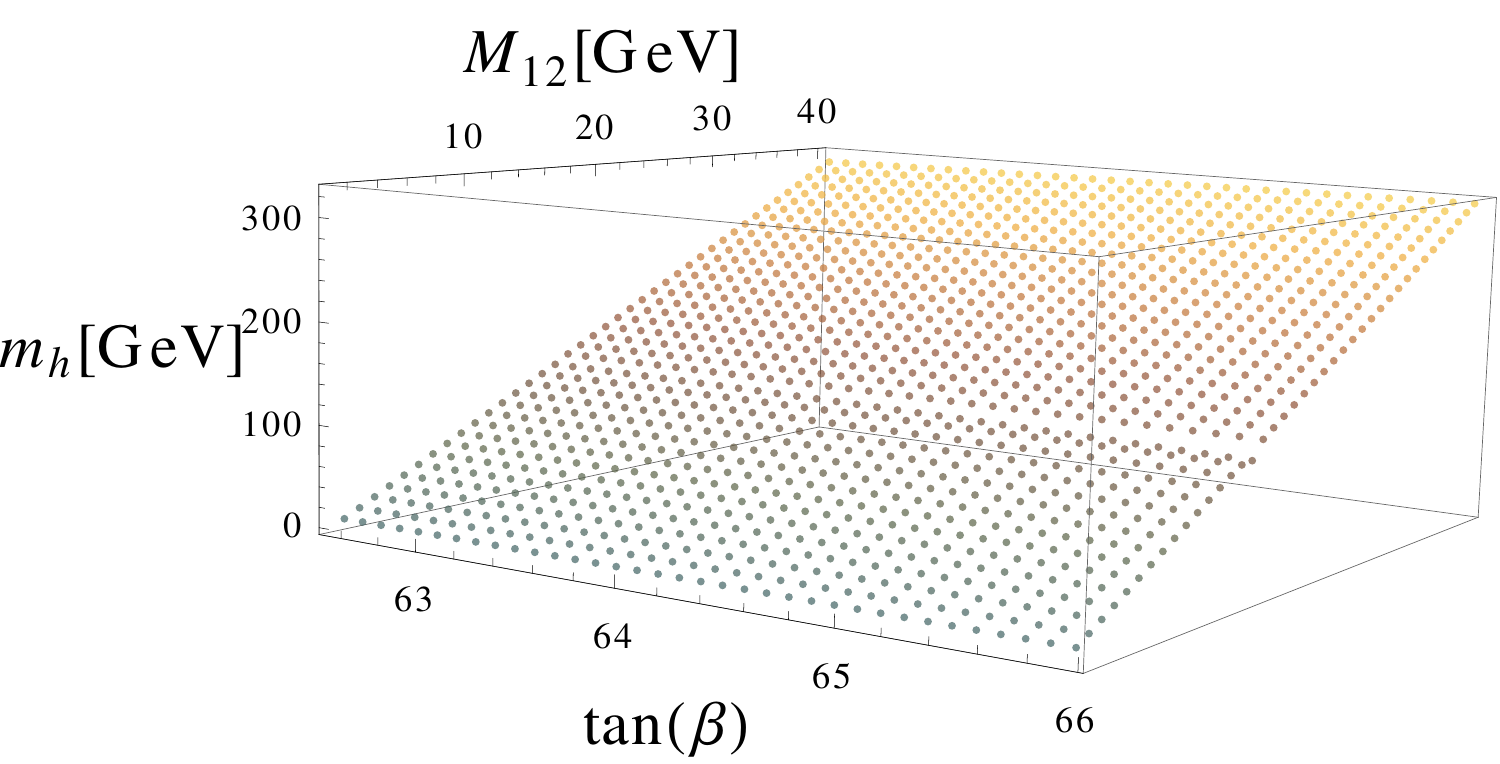}
	}
	\caption{Masses of CP-even neutral scalars $m_h,m_H$ against tan$\beta$ and the soft breaking parameter $M_{12}$.}
	\label{fig8}
\end{figure}

\begin{figure}
	\subfloat{%
		\includegraphics[width=0.49\textwidth]{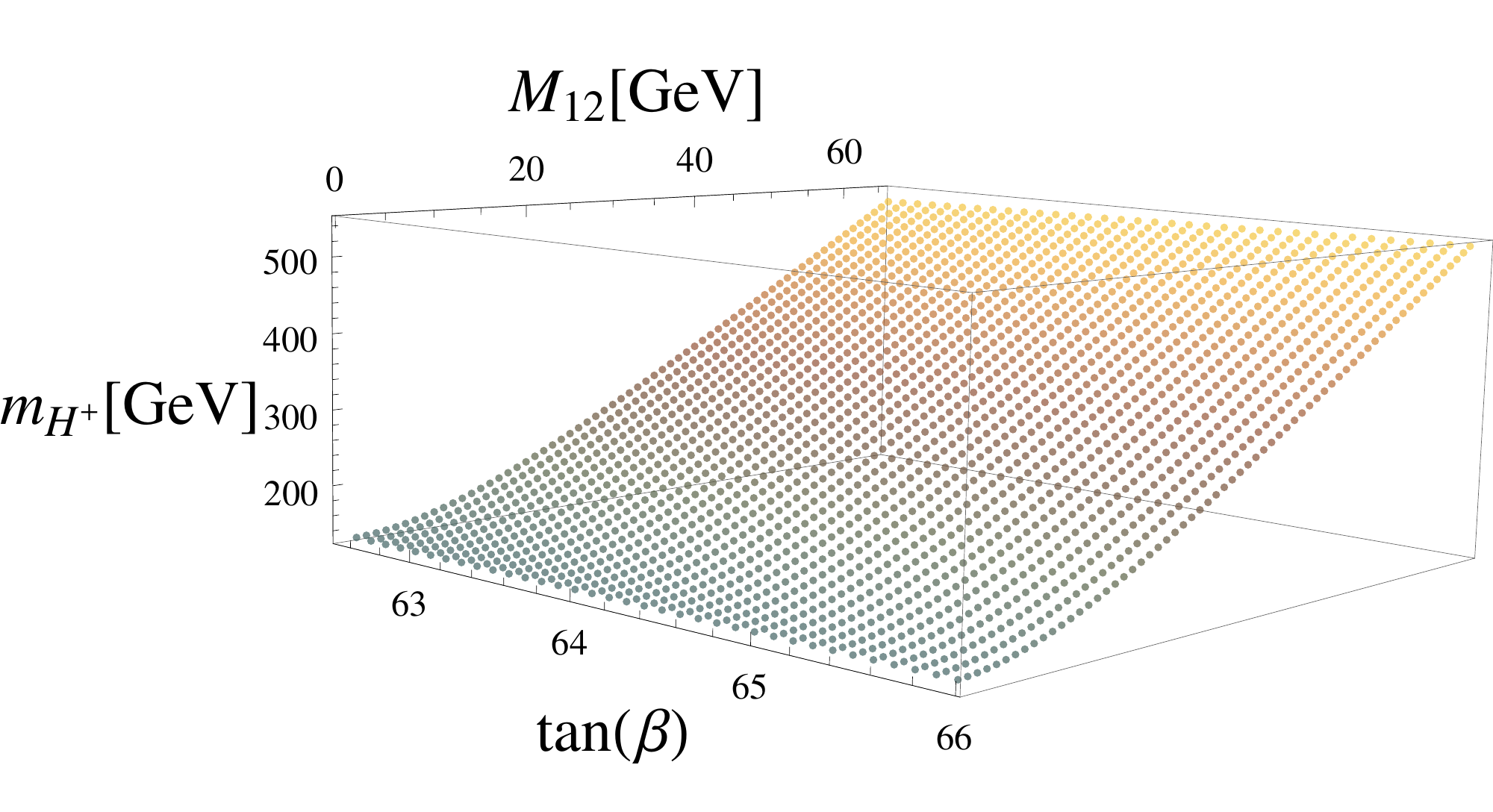}
	}
	\hfill
	\subfloat{%
		\includegraphics[width=0.49\textwidth]{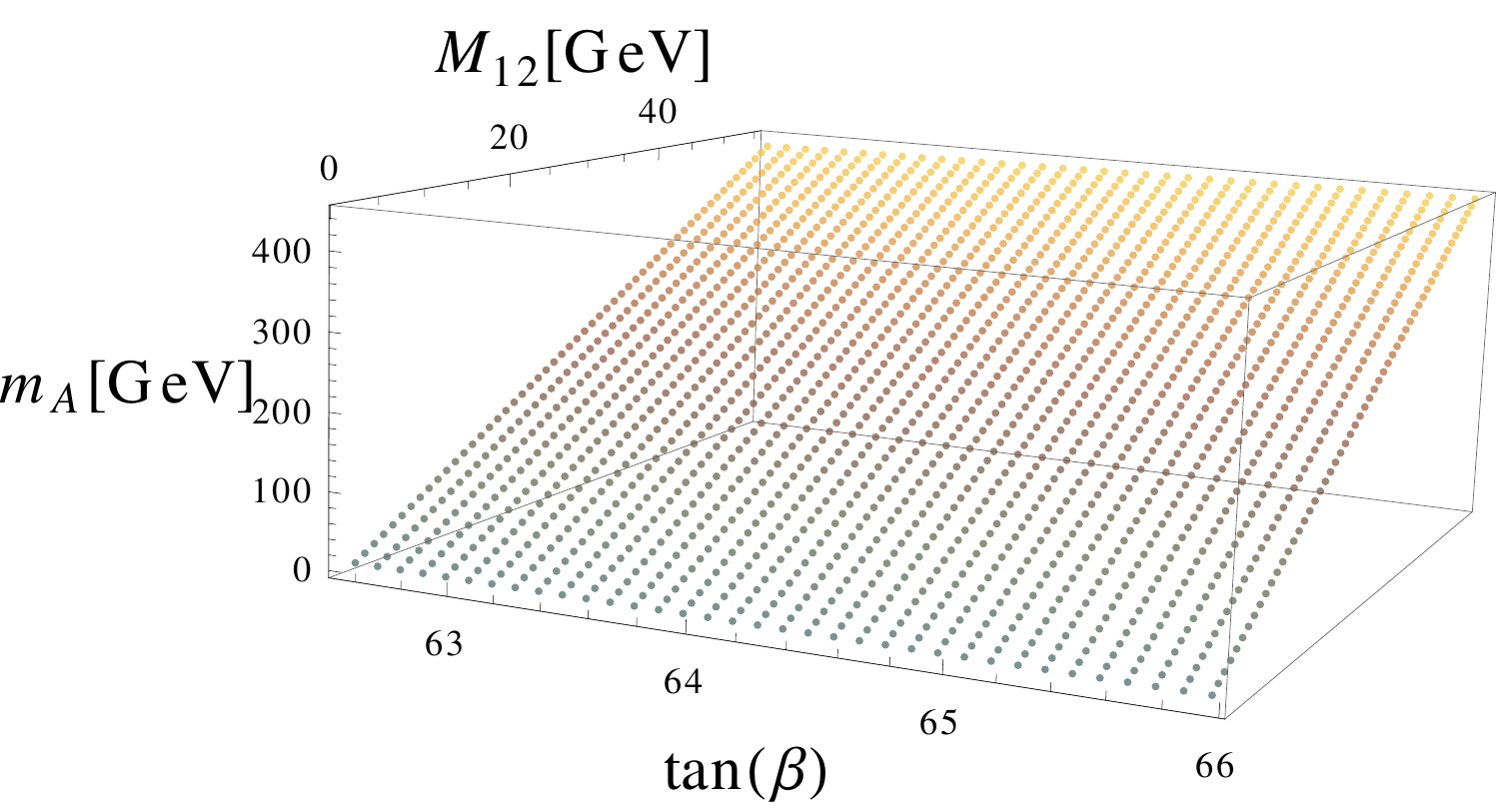}
	}
	\caption{Masses of charged Higgs $H^+$ and pseudoscalar $A$ against tan$\beta$ and the soft breaking parameter $M_{12}$.}
	\label{fig9}
\end{figure}

Compared to $M_{12}$, tan$\beta$ has only a minor impact on the masses. Fig \ref{fig8} shows this for the CP-even eigenstates, with the SM-like $M_{12}$-\textit{independent} eigenstate shown on the left (titled $m_H$ for simplicity), and the $M_{12}$-dependent eigenstate on the right. Figure \ref{fig9} shows the corresponding plots for $m_{H^+}$ (left) and $m_A$ (right). The SM-like Higgs in Fig. \ref{fig8} is the only eigenstate for which there remains a significant difference in stable (top) and unstable (bottom) branch.

A finite $M_{12}$ has a number of implications on the validity of the theory. Most importantly, it has the anticipated effect of allowing the model to produce phenomenologically viable mass spectra by opening up a way to drive $m_h$, $m_A$ and $m_{H^+}$ to higher values. The experimental bounds on the physical Higgs bosons can be translated to a lower bound on $M_{12}$. From the bound of $m_{H^+}$ $>$ 580 GeV \cite{Misiak:2017bgg} in type-Y models it follows that:
\begin{align}
 M_{12} \gtrsim 70 \textrm{ GeV},
 \label{m12bound}
\end{align}
with the exact value depending on tan$\beta$. Re-translated, this condition implies in terms of other boson masses:
\begin{align}
 m_A, m_H &\geq 550 \textrm{ GeV}.
 \label{largehiggsens}
\end{align}
Because of the mixing angle $\alpha$ being close to zero, the SM-like Higgs mass stays almost unchanged. This means that in terms of vacuum stability, the situation also remains consistent with the $\mathbb Z_2$-symmetric case: While there are vacuum-stable solutions to the fixed point equations, only the vacuum-unstable ones include masses around 125 GeV.
The SM-like Higgs is thus in a unique position among the 2HDM bosons, in that its mass cannot be heavily adjusted in this model.

Whereas the SM-like CP-even scalar eigenstate is independent of $M_{12}$, the opposite is true for all other bosons: Even at its minimum, $M_{12}$ $\sim$ 70 GeV is already large enough to make it the controlling factor in generating the masses of the three bosons $H$, $H^+$ and $A$. For larger values of $M_{12}$, the degeneracy in masses becomes even stronger. Therefore, most of the parameter space of viable asymptotically safe 2HDMs falls into the decoupling limit \cite{Gunion:2002zf}, with one SM-like and three heavy bosons with $m_H \approx m_{H^+} \approx m_A \propto M_{12}$.

The high tan$\beta$-values necessary to find fixed points mean that in type II models specifically, bounds from $B_s$ $\rightarrow$ $\mu\mu$ decays are much more restrictive \cite{Cheng:2015yfu, Haller:2018nnx}. They demand heavy boson masses upwards of
\begin{align}
  m_H \approx m_{H^+} \approx m_A &> 3 \textrm{ TeV}.
  \label{largerhiggsens}
\end{align}

\subsection{Stability Analysis}
In order to understand the characteristics of a given fixed point, we study the linearized RG flow around the fixed point, described by the stability matrix given by:
\begin{align}
M_{ij} &= \left. \frac{\partial \beta_{i}}{\partial g_j} \right|_{FP}.
	\end{align} 
In this case, $g_j$ includes gauge, Yukawa and quartic couplings. The number of negative eigenvalues of $M_{ij}$ corresponds to the dimension of the critical surface from which trajectories run into the fixed point. However, it has less significance here: While it is important to confirm that the fixed points are indeed UV-attractive (which they are), both the exact fixed point scale and the low scale initial conditions give additional constraints that intersect non-trivially with the critical surface. The solution to Eq. \eqref{fpe} is always a single trajectory in parameter space. On the other hand, by construction our method of finding fixed points ensures that the solutions found connect to the critical surface.
	
It is therefore necessary to examine which of the initial conditions used is subject to uncertainties, and how these translate to changes in fixed point solutions and thus in Higgs boson mass spectra.

\subsection{Uncertainty Estimates}
\begin{figure}
	\centering
	\includegraphics[width=0.7\textwidth]{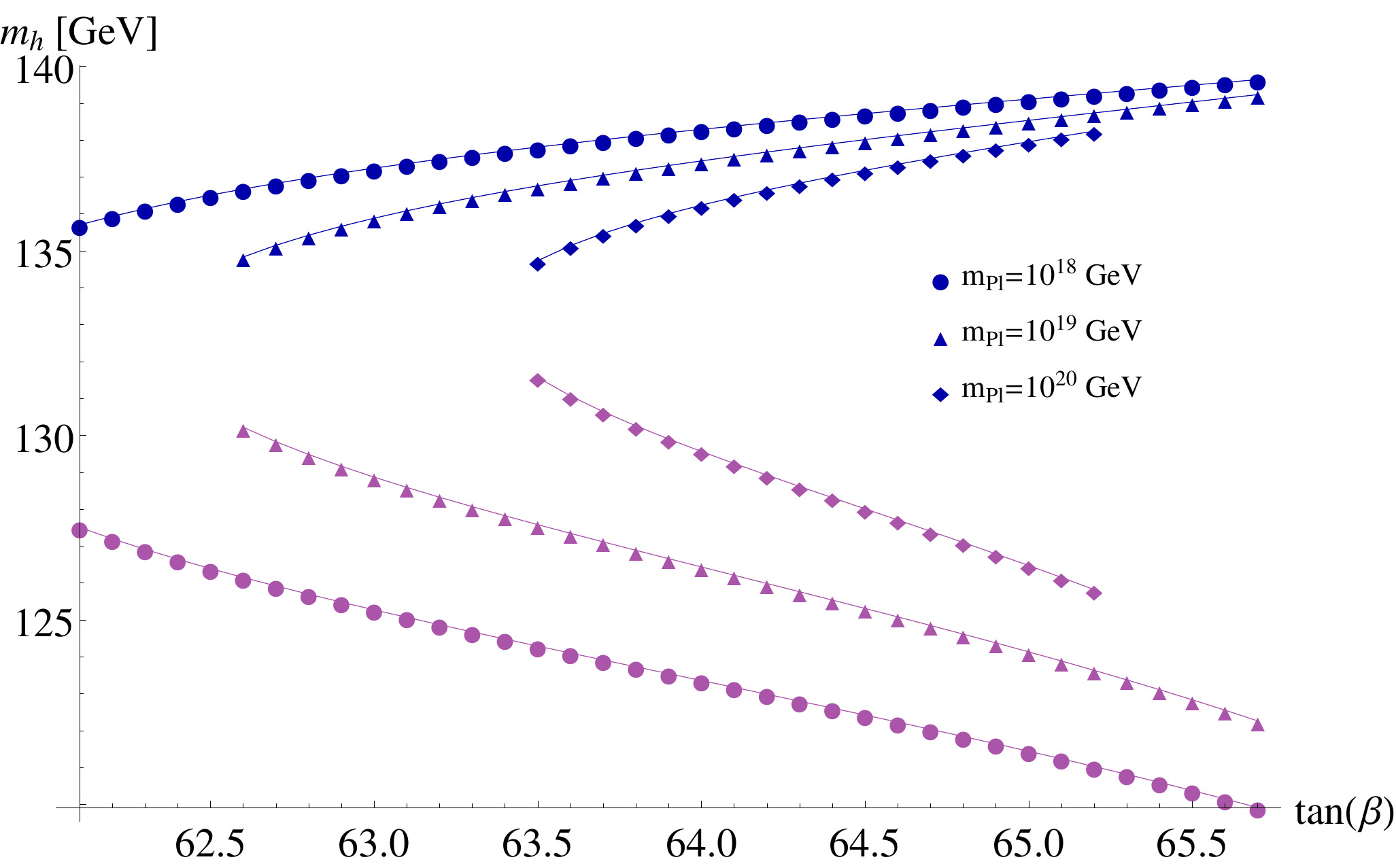}
	
	\caption{Masses of SM-like CP-even scalar Higgs boson, for $M_{12}$ = 380 GeV and fixed point condition set at different scales of $10^{18}$ (blue), $10^{19}$ (violet), and $10^{20}$ (yellow) GeV. The upper branches correspond to the vacuum-stable solutions, the lower branches to the vacuum-unstable ones.}
	\label{fig4}
\end{figure}

\begin{figure}
	\centering
	\subfloat{%
		\includegraphics[width=0.49\textwidth]{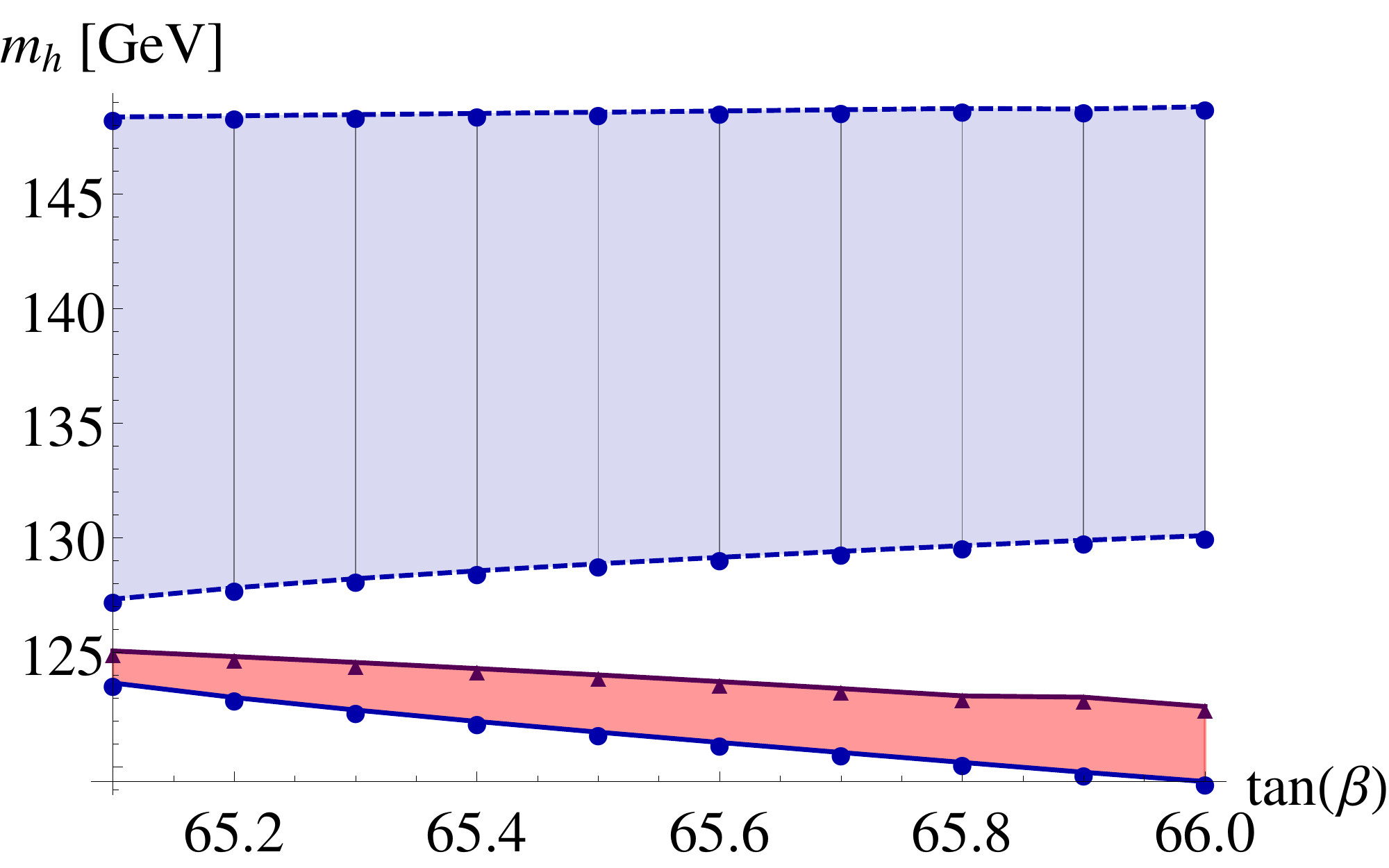}
	}
	\hfill
	\subfloat{%
		\includegraphics[width=0.49\textwidth]{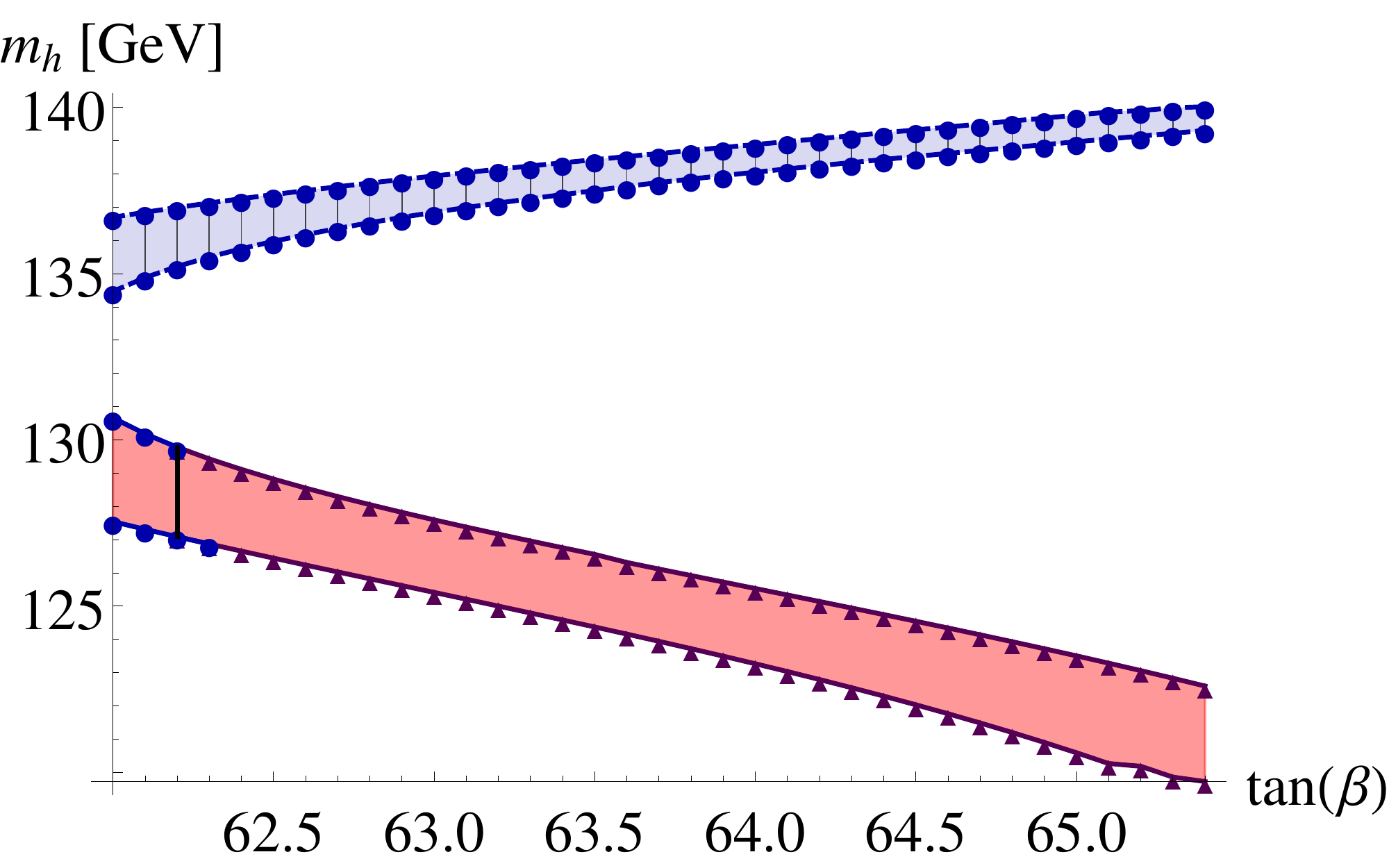}
	}
	\caption{Masses of SM-like CP-even Higgs bosons evaluated at $M_{12}$ $=$ 380 GeV with 1$\sigma$ uncertainty regions from top quark (left) and bottom quark (right) mass initial values of $(160^{+4.8}_{-4.3})$ GeV and $(4.18 \pm 0.03)$ GeV respectively. Stable solutions are marked in blue with dashed outline, unstable solutions in violet.}
	\label{fig5}
\end{figure}

\begin{figure}
	\centering
	
	\subfloat{%
		\includegraphics[width=0.49\textwidth]{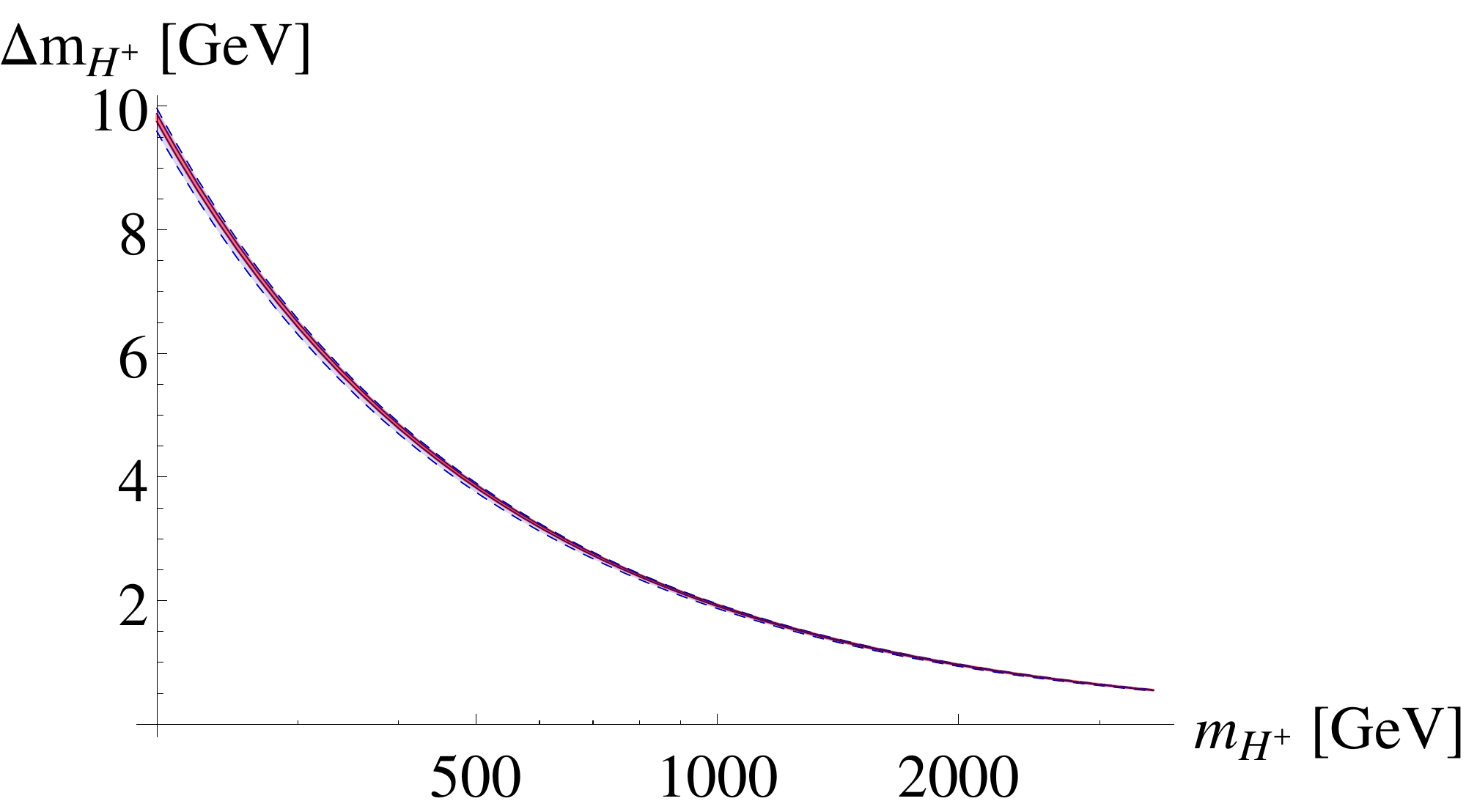}
	}
	\hfill
	\subfloat{%
		\includegraphics[width=0.49\textwidth]{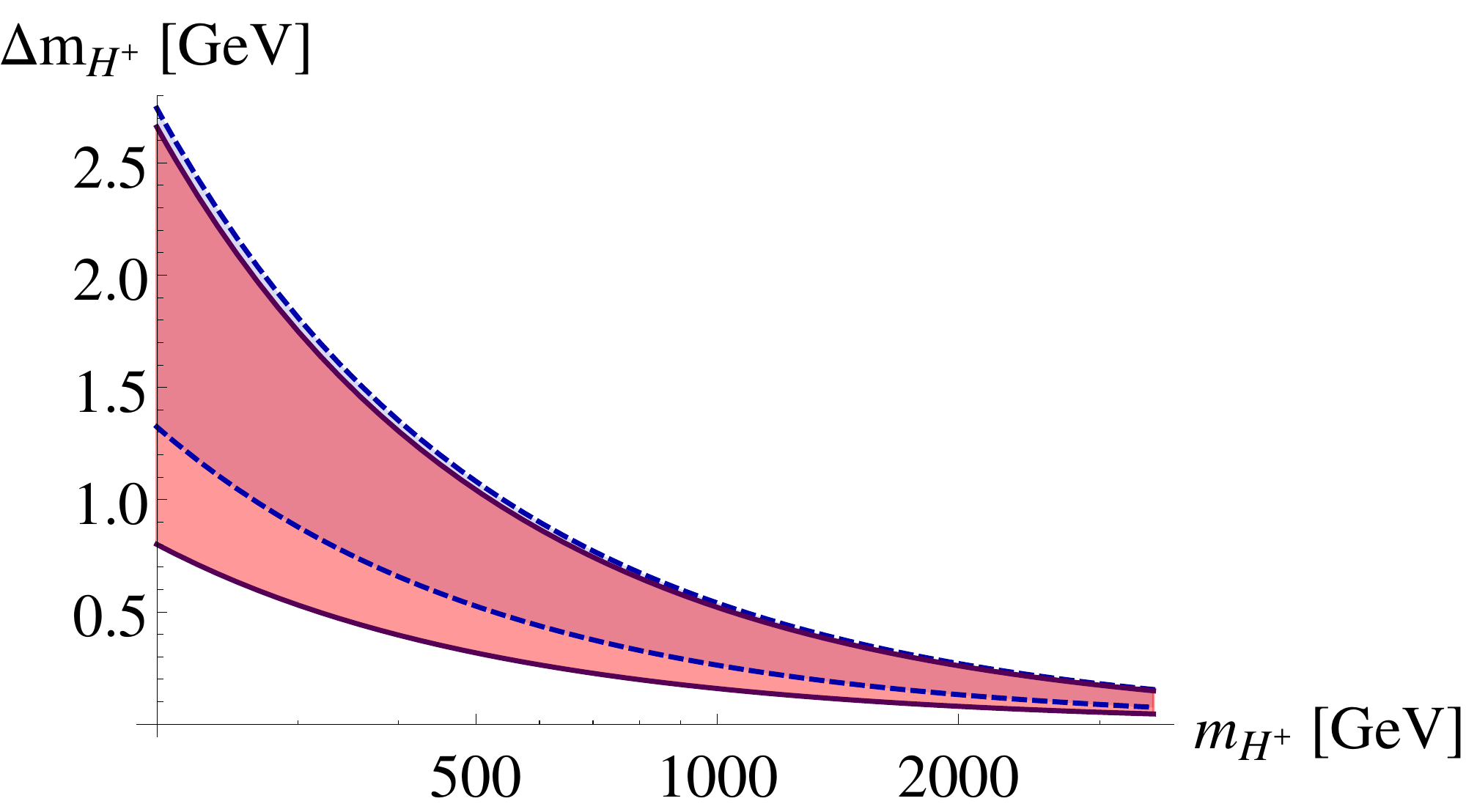}
	}
	\caption{Range of sizes of the uncertainty bands $\Delta m_{H^+}$ on the charged Higgs mass generated by the 1$\sigma$ uncertainty regions from top quark (left) and bottom quark (right) mass initial values of $(160^{+4.8}_{-4.3})$ GeV and $(4.18 \pm 0.03)$ GeV respectively. Stable solutions are marked in blue with dashed lines, unstable solutions in violet.}
	\label{fig5b}
\end{figure}

\begin{figure}
	\centering
	\subfloat{%
		\includegraphics[width=0.49\textwidth]{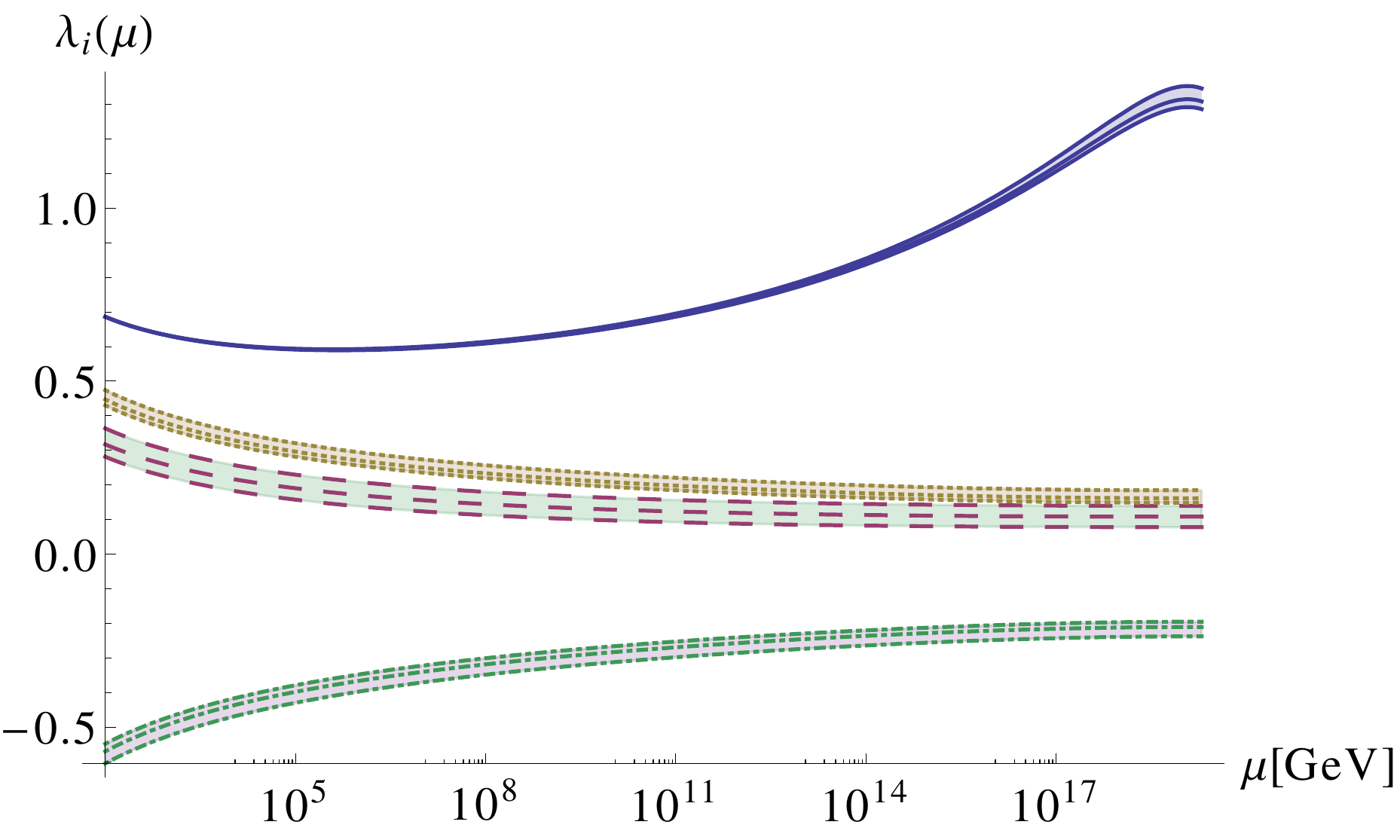}
	}
	\hfill
	\subfloat{%
		\includegraphics[width=0.49\textwidth]{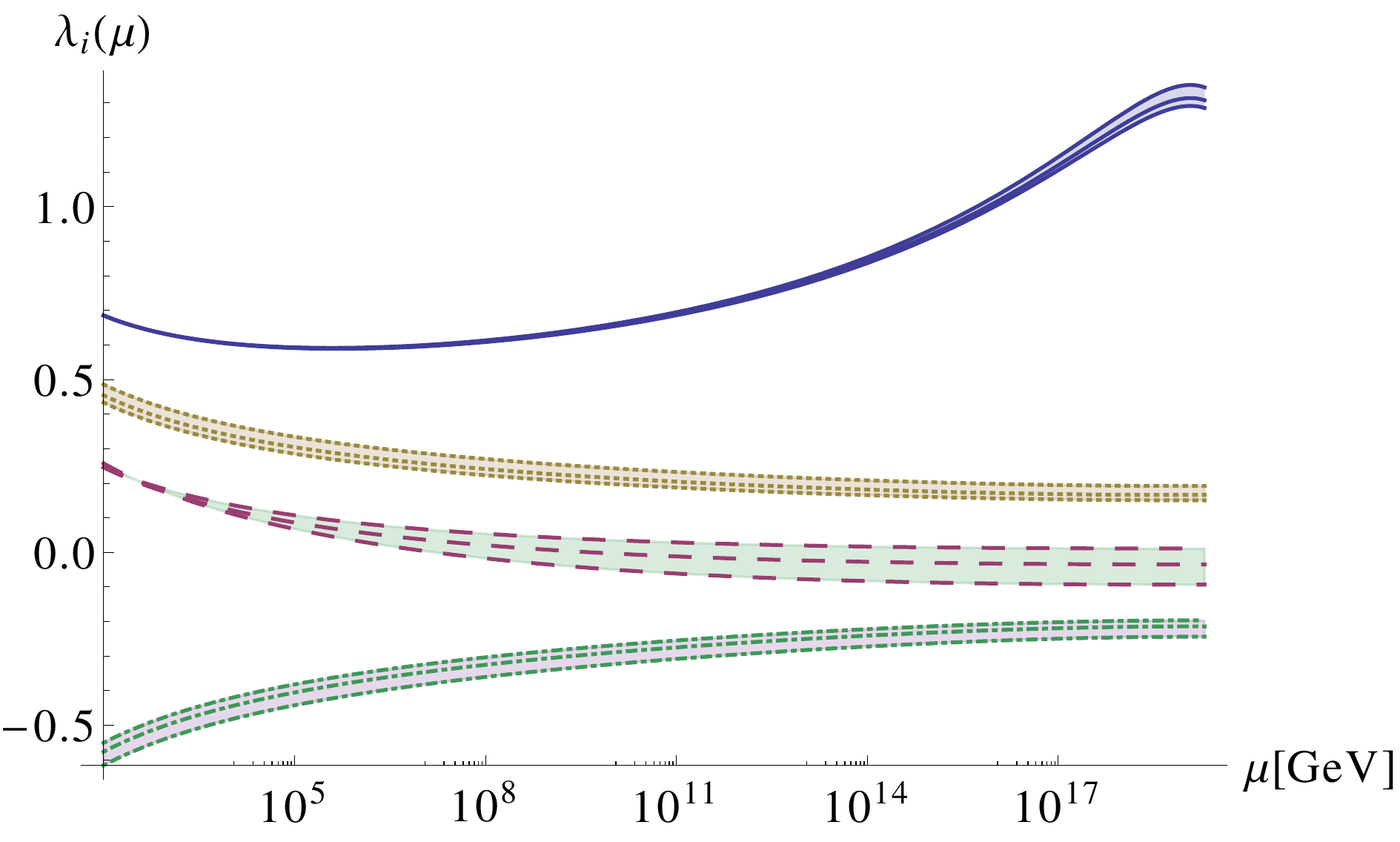}
	}
	\caption{Running quartic couplings with 1$\sigma$ uncertainty intervals from low scale top quark mass initial value for the vacuum-stable (left) and vacuum-unstable (right) fixed point branch. The colours correspond to: $\lambda_1$ (blue), $\lambda_2$ (violet, dashed), $\lambda_3$ (yellow, dotted) and $\lambda_4$ (green, dot-dashed).}
	\label{fig6c}
\end{figure}

\begin{figure}
	\centering
	
	\subfloat{%
		\includegraphics[width=0.49\textwidth]{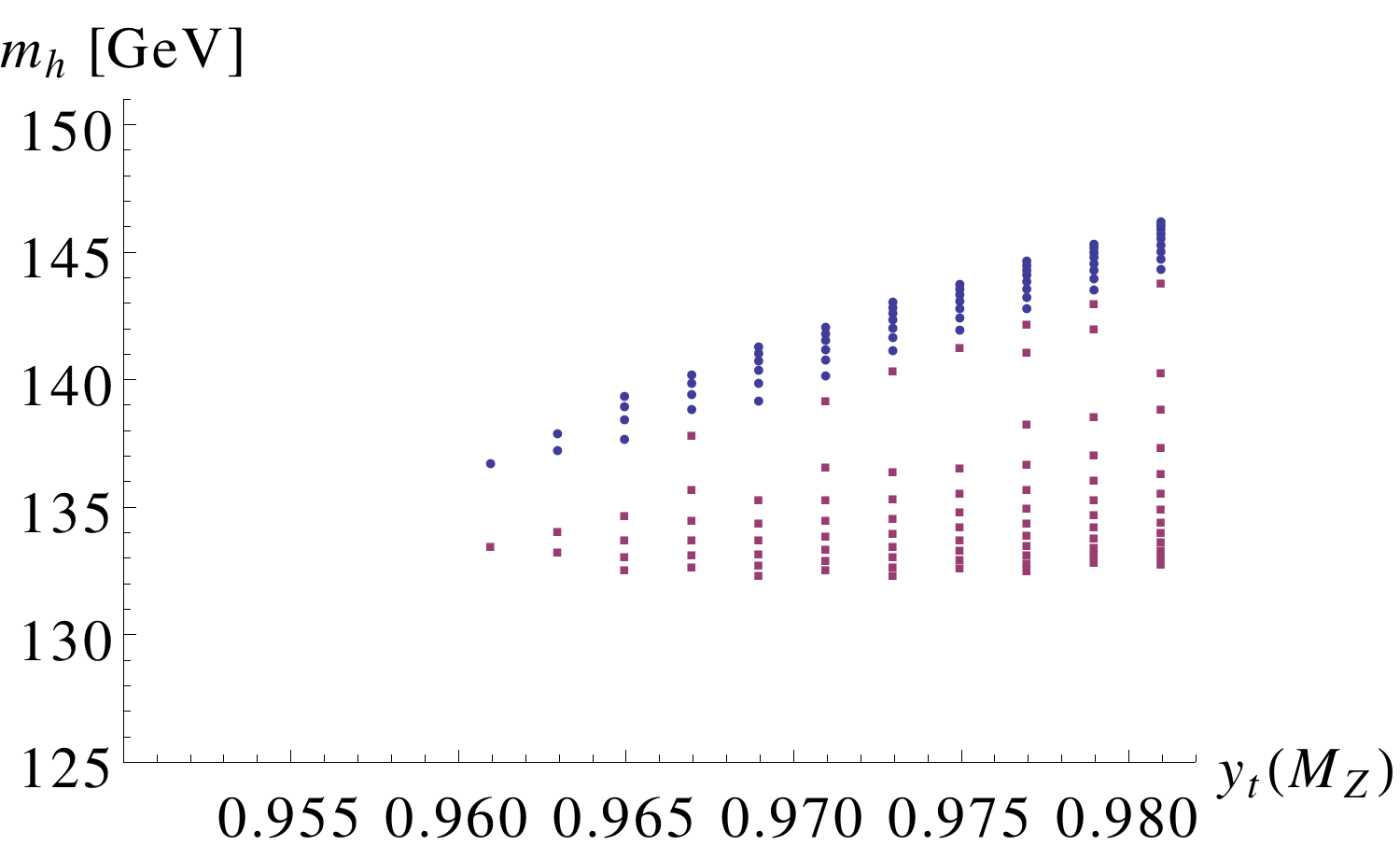}
	}
	\hfill
	\subfloat{%
		\includegraphics[width=0.49\textwidth]{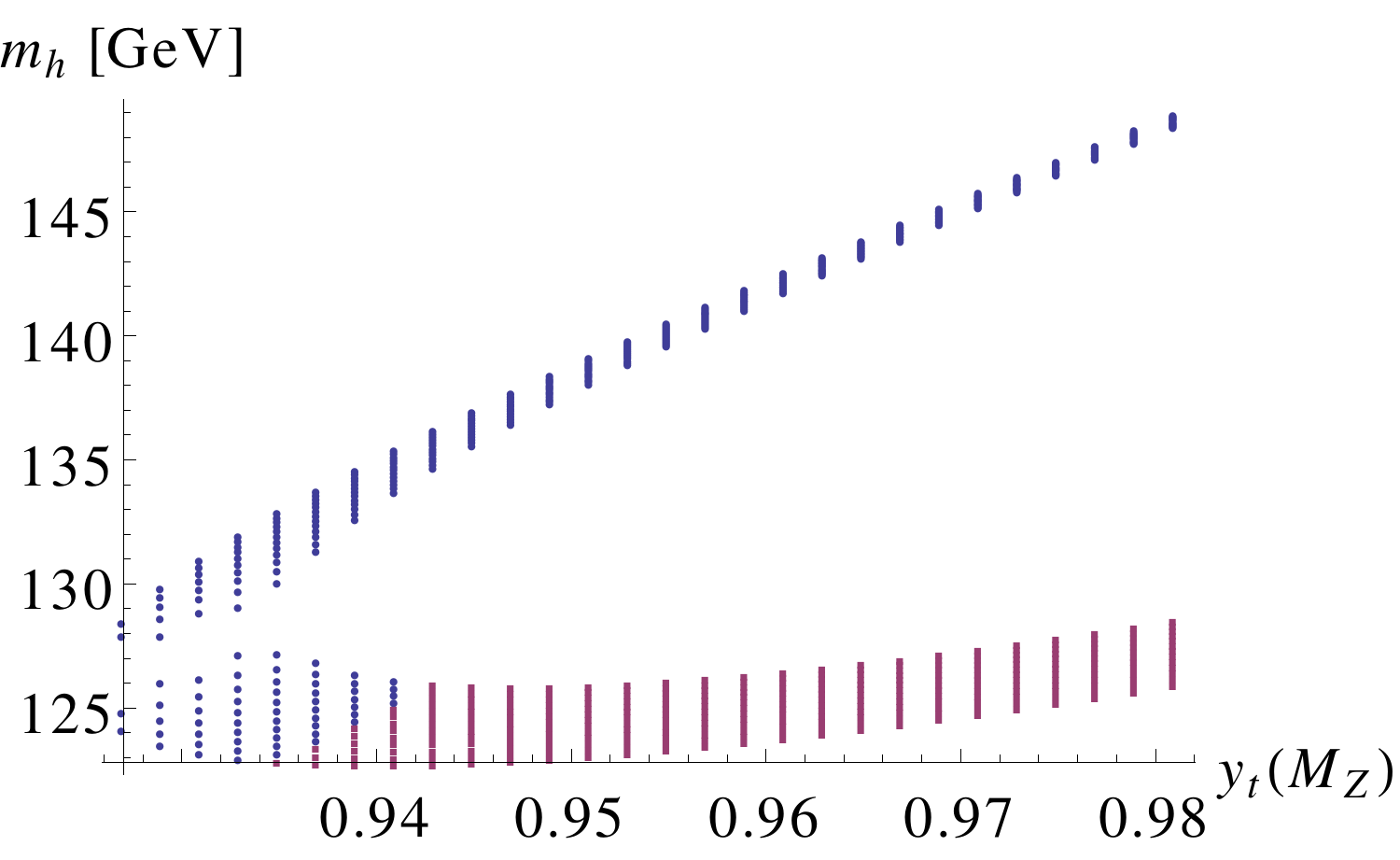}
	}
	\caption{Mass region of the SM-like CP-even scalar Higgs boson from 1$\sigma$ quark mass uncertainties against the top Yukawa initial value $y_t(M_Z)$ for tan$\beta$ $=$ 60 (left) and tan$\beta$ $=$ 64 (right). Stable solutions are marked in blue, unstable solutions in violet.}
	\label{fig7new}
 \end{figure}

There are several factors that influence the fixed point analysis. Below, we look at changes in scale where the fixed point condition is applied, followed by a discussion about low scale top and bottom quark mass uncertainties. Unless stated differently, values for the SM-like CP-even scalar eigenstate (here $h$) will be evaluated using $M_{12,\textrm{Min}}$ $=$ 380 GeV. The pseudoscalar $A$ and $M_{12}$-dependent CP-even scalar eigenstate $H$ are entirely or almost entirely generated by the free parameter $M_{12}$ and therefore have negligible uncertainties. \\
In general, the models are studied with the condition that the quartic coupling beta functions become zero at $m_{Pl}$ $=$ $1.2\cdot10^{19}$ GeV. The mass spectrum shows a minor dependence on where exactly the fixed points are assumed to occur. The masses $m_h$ are shown in Figure \ref{fig4} for different fixed point scales. We see that lower fixed point scales correspond to a larger difference between vacuum-stable (upper) and vacuum-unstable (lower) branches, and notably bring down the lower branch mass values. Also, the tan$\beta$-range in which fixed points can be found changes with the fixed point scale: When the fixed points scale is chosen at higher values than $m_{Pl}$, the divergence of large Yukawa couplings, especially $y_b$ becomes an even more pronounced problem. \\
The mass spectra depend on the initial values chosen for the gauge and Yukawa couplings. The dependency on the top quark mass turns out to be especially strong. In Figure \ref{fig5} the SM-like Higgs boson masses are shown for the $1\sigma$ deviation bands of the $\overline{MS}$ top and bottom quark masses of $m_t(m_t)$ $=$ $(160^{+4.8}_{-4.3})$ GeV and $m_b(m_b)$ $=$ $(4.18 \pm 0.03)$ GeV \cite{Agashe:2014kda}. For the charged Higgs mass, the uncertainty generally grows with tan$\beta$, but also depends on the soft breaking parameter, as the quartic coupling contributions to the boson mass weaken with $\sqrt{M_{12}}^{-1}$. The ranges of uncertanties $\Delta m_{H^+}$ against the central mass value $m_{H^+}$ are shown in Figure \ref{fig5b}. It has to be noted that the tan$\beta$-interval shown in the first (left) plot of Figure \ref{fig5} is smaller than the intervals shown in the corresponding right hand plot, or in Figures \ref{fig3}-\ref{figalphabeta}: The reason is that different Yukawa initial values not only change the mass spectrum, but also the region in parameter space for which a fixed point exists: On one hand, smaller quark masses mean that the Yukawa couplings require larger tan$\beta$ to fulfil the fixed point condition. On the other hand, larger quark masses move the Landau pole of the Yukawa couplings to lower scales. The first effect is very noticeable for lower values of $m_t$. For the same reason, the variance in uncertainties at fixed values of $M_{12}$ is much smaller in the first plot of Figure \ref{fig5b} compared to the second.\\
Figure \ref{fig6c} shows a typical set of running quartic couplings each for the vacuum-stable and unstable branch and how different top quark mass initial values influence $\lambda_i$. For most couplings, the $m_t$-induced relative uncertainty becomes smaller as the scale decreases, even if the coupling itself becomes larger. This is most notable for $\lambda_1$ (blue) and $\lambda_2$ (violet). The quartic coupling $\lambda_5$ is not shown in these graphs, as $\lambda_5$ $=$ $0$ is an exact solution of the fixed point equations regardless of initial conditions. A last important detail to take note of is the range of $\lambda_2$ in the right graph: Depending on the exact initial conditions, $\lambda_2$ may become negative (and thereby break the vacuum stability conditions of Eq. \eqref{stabilitycond}) as early as $10^8$ GeV, or not at all. 
To further analyse the influence of the exact starting parameters on the Higgs mass, all SM-like Higgs mass values generated by a fixed point solution can be shown against the corresponding top quark Yukawa initial value. Figure \ref{fig7new} shows the Higgs mass $m_h$ for the 1$\sigma$ regions of $m_b$ and $m_t$ for two different values of tan$\beta$. The vertical spread in points is generated by shifting the bottom quark initial value. Once again, all vacuum-stable points are coloured blue, the vacuum-instable ones violet. 
The right hand graph in particular illustrates well the two branches of fixed point solutions: The branch corresponding to higher Higgs boson masses is fully vacuum-stable, whereas the lower mass branch can only be vacuum-stable for top quark mass values on the lower end of its 1$\sigma$ band. Comparing both graphs also shows how tan$\beta$ needs to be of certain size to facilitate the existence of fixed points. For the central mass values of $m_b$ and $m_t$, there are no fixed points at tan$\beta$ $=$ 60. However, as shown in the left graph, fixed points can be found there if either or both initial values are slightly larger.

\section{Summary}
Proposing simultaneously vanishing quartic coupling beta functions at the Planck scale severely constrains the 2HDM parameter space, but is possible in a way similar to the SM. As such, the 2HDM likewise supports the idea of being extended to high scales through means of asymptotic safety.

The parameter tan$\beta$ needs to be large, as both the up-type and the down-type Yukawa couplings have to be large in order to keep the positive contributions in quartic couplings beta functions in check. For the same reason, only type II and type Y models are viable, while type I and type X models are not. In the type II/type Y models studied, there always exists a tan$\beta$-interval in which fixed points can be found, see Eq. \eqref{interval}. The most minimal model that also agrees with all experimental bounds is the softly-broken $\mathbb{Z}_2$-symmetric 2HDM, cf. Figs. \ref{lc4} - \ref{fig9}.

The allowed parameter region defined by the fixed point assumption meets the characteristics of the decoupled alignment limit, with three heavy Higgs bosons $m_H$ $\approx$ $m_A$ $\approx$ $m_{H^+}$ $\propto$ $M_{12}$, and $|\beta-\alpha|$ $\approx$ $\frac{\pi}{2}$. To be consistent with experimental constraints, a lower bound is given on $M_{12}$ $>$ 70 GeV (380 Gev) for type Y (type II) models, corresponding to the charged Higgs limits for type Y. This implies lower limits on $m_A$ and $m_H$, see Eqs. \ref{largehiggsens} and \ref{largerhiggsens}.

Similar to the SM, both the existence of fixed points and the vacuum stability depend strongly on the low scale initial values, most notably the exact top quark mass. As illustrated in Fig. \ref{fig7new}, the central values of Higgs and top quark mass indicate an instable vacuum, but are rather close to the criticality border. Fixed point solutions with stability can be possible by having the $\overline{\textrm{MS}}$ top quark mass values lower than 160 GeV \cite{Agashe:2014kda} or the fixed point scale set below $m_{Pl}$ $=$ $1.2\cdot10^{19}$ GeV. A more precise determination of $m_t$ and $m_h$ by future experiments will allow a more definite statement.   

As it stands, the 2HDM does not solve the stability problem of the SM. Instead, the situation is mirrored, or even worse.\\
\textbf{Note added:} During the final phase of this project, an analysis of 2HDM fixed points using slightly different methodology appeared \cite{McDowall:2018ulq}. While our approach differs in details, we agree with the general conclusion that fixed points in type II 2HDMs are possible.

\section*{Acknowledgements}
We are happy to thank Gudrun Hiller for fruitful discussions, and Otto Eberhardt for helpful comments.

\clearpage

\appendix
\section{$\beta$ functions}
In this section, the $\beta$ functions $\beta_{g_i}$ $=$ $\frac{dg_i(\mu)}{d \log(\mu)}$ $=$ $\mu \frac{dg_i(\mu)}{d\mu}$ are listed on 2-loop level for the most general model used (i.e. the softly-broken $\mathbb{Z}_2$-symmetric 2HDM type II, which means that $\lambda_6$ and $\lambda_7$ do not appear). They were calculated with the Mathematica package SARAH \cite{Staub:2008uz, Staub:2013tta}. The general procedure of how to derive 2-loop RGEs for general field theories has been outlined in \cite{Machacek:1983tz, Machacek:1983fi, Machacek:1984zw}.\\
In the case of Yukawa couplings, first and second generation contributions have been neglected. The Yukawa matrices have therefore been restricted to their respective $(3,3)$-entries.

The $\beta$ functions for the gauge couplings are given by:

\begin{align}
 16 \pi^2 \beta_{g_1} &= 7 \text{$g_1$}^3 + \frac{1}{288 \pi ^2}\Big(\text{$g_1$}^3 \Big(208 \text{$g_1$}^2+3 \Big(36 \text{$g_2$}^2+88 \text{$g_3$}^2-5 \text{$\lambda_b$}^2-15 \text{$\lambda_\tau$}^2-17 \text{$\lambda_t$}^2\Big)\Big)\Big),\\
 16 \pi^2 \beta_{g_2} &= - 3\text{$g_2$}^3 +  \frac{1}{32 \pi ^2}\Big(-\text{$g_2$}^3 \Big(-4 \text{$g_1$}^2-16 \text{$g_2$}^2-24 \text{$g_3$}^2+3 \text{$\lambda_b$}^2+\text{$\lambda_\tau$}^2+3 \text{$\lambda_t$}^2\Big)\Big),\\
 16 \pi^2 \beta_{g_3} &=  - 7\text{$g_3$}^3 + \frac{1}{96 \pi ^2}\Big(-\text{$g_3$}^3 \Big(3 \Big(4 \Big(13 \text{$g_3$}^2+\text{$\lambda_b$}^2+\text{$\lambda_t$}^2\Big)-9 \text{$g_2$}^2\Big)-11 \text{$g_1$}^2\Big)\Big).
\end{align}

The $\beta$ functions for the quartic Higgs couplings $\lambda_i$ for the softly-broken type II 2HDM are given by:
\begin{align}
 16\pi^2\beta_{\lambda_1} &= \frac{1}{4} \Big(6 \text{$g_1$}^2 \Big(\text{$g_2$}^2-2 \text{$\lambda_1$}\Big)+3 \text{$g_1$}^4-36 \text{$g_2$}^2 \text{$\lambda_1$}+9 \text{$g_2$}^4\nonumber \\
  &+8 \Big(2 \text{$\lambda_1$} \Big(3 \text{$\lambda_b$}^2+\text{$\lambda_\tau$}^2\Big)+6 \text{$\lambda_1$}^2+2 \text{$\lambda_3$} \text{$\lambda_4$}+2 \text{$\lambda_3$}^2+\text{$\lambda_4$}^2+\text{$\lambda_5$}^2-6 \text{$\lambda_b$}^4-2 \text{$\lambda_\tau$}^4\Big)\Big) \nonumber \\
  &+ \frac{1}{384 \pi ^2}\Big[\text{$g_1$}^2 \Big(6 \text{$g_2$}^2 \Big(39 \text{$\lambda_1$}+20 \text{$\lambda_4$}+36 \text{$\lambda_b$}^2+44 \text{$\lambda_\tau$}^2\Big)-303 \text{$g_2$}^4\nonumber \\
  &+4 \Big(25 \text{$\lambda_1$} \Big(\text{$\lambda_b$}^2+3 \text{$\lambda_\tau$}^2\Big)+108 \text{$\lambda_1$}^2+4 \Big(6 \Big(2 \text{$\lambda_3$} \text{$\lambda_4$}+2 \text{$\lambda_3$}^2+\text{$\lambda_4$}^2\Big)\nonumber \\
  &-3 \text{$\lambda_5$}^2+4 \text{$\lambda_b$}^4-12 \text{$\lambda_\tau$}^4\Big)\Big)\Big)+\text{$g_1$}^4 \Big(-573 \text{$g_2$}^2+651 \text{$\lambda_1$}+60 \Big(2 \text{$\lambda_3$}+\text{$\lambda_4$}+\text{$\lambda_b$}^2-5 \text{$\lambda_\tau$}^2\Big)\Big)\nonumber \\
  &-393 \text{$g_1$}^6+3 \Big(-3 \text{$g_2$}^4 \Big(17 \text{$\lambda_1$}+4 \Big(-10 \text{$\lambda_3$}-5 \text{$\lambda_4$}+3 \text{$\lambda_b$}^2+\text{$\lambda_\tau$}^2\Big)\Big)\nonumber \\
  &+12 \text{$g_2$}^2 \Big(5 \text{$\lambda_1$} \Big(3 \text{$\lambda_b$}^2+\text{$\lambda_\tau$}^2\Big)+36 \text{$\lambda_1$}^2+4 (2 \text{$\lambda_3$}+\text{$\lambda_4$})^2\Big)+291 \text{$g_2$}^6\nonumber \\
  &-8 \Big(\text{$\lambda_1$} \Big(-80 \text{$g_3$}^2 \text{$\lambda_b$}^2+20 \text{$\lambda_3$} \text{$\lambda_4$}+20 \text{$\lambda_3$}^2+12 \text{$\lambda_4$}^2+14 \text{$\lambda_5$}^2+3 \text{$\lambda_b$}^4+\text{$\lambda_\tau$}^4\Big)\nonumber \\
  &+4 \Big(16 \text{$g_3$}^2 \text{$\lambda_b$}^4+\text{$\lambda_5$}^2 (10 \text{$\lambda_3$}+11 \text{$\lambda_4$})+8 \text{$\lambda_3$} \text{$\lambda_4$}^2+6 \text{$\lambda_3$}^2 \text{$\lambda_4$}\nonumber \\
  &+4 \text{$\lambda_3$}^3+3 \text{$\lambda_4$}^3-5 \Big(3 \text{$\lambda_b$}^6+\text{$\lambda_\tau$}^6\Big)\Big)+24 \text{$\lambda_1$}^2 \Big(3 \text{$\lambda_b$}^2+\text{$\lambda_\tau$}^2\Big)+78 \text{$\lambda_1$}^3\Big)\nonumber \\
  &-24 \text{$\lambda_t$}^2 \Big(3 \text{$\lambda_1$} \text{$\lambda_b$}^2+8 \text{$\lambda_3$} \text{$\lambda_4$}+8 \text{$\lambda_3$}^2+4 \Big(\text{$\lambda_4$}^2+\text{$\lambda_5$}^2\Big)-4 \text{$\lambda_b$}^4\Big)\Big)\Big].
\end{align}

\begin{align}
 16\pi^2\beta_{\lambda_2} &= \frac{1}{4} \Big(6 \text{$g_1$}^2 \Big(\text{$g_2$}^2-2 \text{$\lambda_2$}\Big)+3 \text{$g_1$}^4-36 \text{$g_2$}^2 \text{$\lambda_2$}+9 \text{$g_2$}^4\nonumber \\
  &+8 \Big(6 \text{$\lambda_2$} \text{$\lambda_t$}^2+6 \text{$\lambda_2$}^2+2 \text{$\lambda_3$} \text{$\lambda_4$}+2 \text{$\lambda_3$}^2+\text{$\lambda_4$}^2+\text{$\lambda_5$}^2-6 \text{$\lambda_t$}^4\Big)\Big) \nonumber \\
 &+ \frac{1}{384 \pi ^2}\Big[\text{$g_1$}^2 \Big(6 \text{$g_2$}^2 \Big(39 \text{$\lambda_2$}+20 \text{$\lambda_4$}+84 \text{$\lambda_t$}^2\Big)-303 \text{$g_2$}^4\nonumber \\
  &+48 \Big(9 \text{$\lambda_2$}^2+4 \text{$\lambda_3$} \text{$\lambda_4$}+4 \text{$\lambda_3$}^2+2 \text{$\lambda_4$}^2-\text{$\lambda_5$}^2\Big)+340 \text{$\lambda_2$} \text{$\lambda_t$}^2-128 \text{$\lambda_t$}^4\Big)\nonumber \\
  &+3 \text{$g_1$}^4 \Big(-191 \text{$g_2$}^2+217 \text{$\lambda_2$}+40 \text{$\lambda_3$}+20 \text{$\lambda_4$}-76 \text{$\lambda_t$}^2\Big)-393 \text{$g_1$}^6\nonumber \\
  &+3 \Big(\text{$g_2$}^4 \Big(-51 \text{$\lambda_2$}+60 (2 \text{$\lambda_3$}+\text{$\lambda_4$})-36 \text{$\lambda_t$}^2\Big)\nonumber \\
  &+12 \text{$g_2$}^2 \Big(4 \Big(9 \text{$\lambda_2$}^2+(2 \text{$\lambda_3$}+\text{$\lambda_4$})^2\Big)+15 \text{$\lambda_2$} \text{$\lambda_t$}^2\Big)+291 \text{$g_2$}^6\nonumber \\
  &+8 \Big(\text{$\lambda_t$}^4 \Big(-64 \text{$g_3$}^2-3 \text{$\lambda_2$}+12 \text{$\lambda_b$}^2\Big)+\text{$\lambda_2$} \text{$\lambda_t$}^2 \Big(80 \text{$g_3$}^2-9 \Big(8 \text{$\lambda_2$}+\text{$\lambda_b$}^2\Big)\Big)\nonumber \\
  &-2 \text{$\lambda_5$}^2 (7 \text{$\lambda_2$}+20 \text{$\lambda_3$}+22 \text{$\lambda_4$})-2 \Big(2 \text{$\lambda_4$}^2 (3 \text{$\lambda_2$}+8 \text{$\lambda_3$})+2 \text{$\lambda_3$} \text{$\lambda_4$} (5 \text{$\lambda_2$}+6 \text{$\lambda_3$})\nonumber \\
  &+10 \text{$\lambda_2$} \text{$\lambda_3$}^2+39 \text{$\lambda_2$}^3+8 \text{$\lambda_3$}^3+6 \text{$\lambda_4$}^3\Big)-12 \text{$\lambda_b$}^2 \Big(2 \text{$\lambda_3$} \text{$\lambda_4$}+2 \text{$\lambda_3$}^2+\text{$\lambda_4$}^2+\text{$\lambda_5$}^2\Big)\nonumber \\
  &-4 \text{$\lambda_\tau$}^2 \Big(2 \text{$\lambda_3$} \text{$\lambda_4$}+2 \text{$\lambda_3$}^2+\text{$\lambda_4$}^2+\text{$\lambda_5$}^2\Big)+60 \text{$\lambda_t$}^6\Big)\Big)\Big].
\end{align}

\begin{align}
 16\pi^2\beta_{\lambda_3} &= \frac{1}{4} \Big(-6 \text{$g_1$}^2 \Big(\text{$g_2$}^2+2 \text{$\lambda_3$}\Big)+3 \text{$g_1$}^4-36 \text{$g_2$}^2 \text{$\lambda_3$}+9 \text{$g_2$}^4\nonumber \\
  &+8 \Big(\text{$\lambda_1$} (3 \text{$\lambda_3$}+\text{$\lambda_4$})+\text{$\lambda_2$} (3 \text{$\lambda_3$}+\text{$\lambda_4$})+\text{$\lambda_3$} \Big(2 \text{$\lambda_3$}+3 \text{$\lambda_b$}^2+\text{$\lambda_\tau$}^2\Big)\nonumber \\
  &+3 \text{$\lambda_t$}^2 \Big(\text{$\lambda_3$}-2 \text{$\lambda_b$}^2\Big)+\text{$\lambda_4$}^2+\text{$\lambda_5$}^2\Big)\Big) \nonumber \\
 &+ \frac{1}{384 \pi ^2}\Big[\text{$g_1$}^2 \Big(6 \text{$g_2$}^2 \Big(39 \text{$\lambda_1$}+20 \text{$\lambda_4$}+36 \text{$\lambda_b$}^2+44 \text{$\lambda_\tau$}^2\Big)-303 \text{$g_2$}^4\nonumber \\
  &+4 \Big(25 \text{$\lambda_1$} \Big(\text{$\lambda_b$}^2+3 \text{$\lambda_\tau$}^2\Big)+108 \text{$\lambda_1$}^2+4 \Big(6 \Big(2 \text{$\lambda_3$} \text{$\lambda_4$}+2 \text{$\lambda_3$}^2+\text{$\lambda_4$}^2\Big)-3 \text{$\lambda_5$}^2+4 \text{$\lambda_b$}^4-12 \text{$\lambda_\tau$}^4\Big)\Big)\Big)\nonumber \\
  &+\text{$g_1$}^4 \Big(-573 \text{$g_2$}^2+651 \text{$\lambda_1$}+60 \Big(2 \text{$\lambda_3$}+\text{$\lambda_4$}+\text{$\lambda_b$}^2-5 \text{$\lambda_\tau$}^2\Big)\Big)-393 \text{$g_1$}^6\nonumber \\
  &+3 \Big(-3 \text{$g_2$}^4 \Big(17 \text{$\lambda_1$}+4 \Big(-10 \text{$\lambda_3$}-5 \text{$\lambda_4$}+3 \text{$\lambda_b$}^2+\text{$\lambda_\tau$}^2\Big)\Big)\nonumber \\
  &+12 \text{$g_2$}^2 \Big(5 \text{$\lambda_1$} \Big(3 \text{$\lambda_b$}^2+\text{$\lambda_\tau$}^2\Big)+36 \text{$\lambda_1$}^2+4 (2 \text{$\lambda_3$}+\text{$\lambda_4$})^2\Big)+291 \text{$g_2$}^6\nonumber \\
  &-8 \Big(\text{$\lambda_1$} \Big(-80 \text{$g_3$}^2 \text{$\lambda_b$}^2+20 \text{$\lambda_3$} \text{$\lambda_4$}+20 \text{$\lambda_3$}^2+12 \text{$\lambda_4$}^2+14 \text{$\lambda_5$}^2+3 \text{$\lambda_b$}^4+\text{$\lambda_\tau$}^4\Big)\nonumber \\
  &+4 \Big(16 \text{$g_3$}^2 \text{$\lambda_b$}^4+\text{$\lambda_5$}^2 (10 \text{$\lambda_3$}+11 \text{$\lambda_4$})+8 \text{$\lambda_3$} \text{$\lambda_4$}^2+6 \text{$\lambda_3$}^2 \text{$\lambda_4$}+4 \text{$\lambda_3$}^3+3 \text{$\lambda_4$}^3-5 \Big(3 \text{$\lambda_b$}^6+\text{$\lambda_\tau$}^6\Big)\Big)\nonumber \\
  &+24 \text{$\lambda_1$}^2 \Big(3 \text{$\lambda_b$}^2+\text{$\lambda_\tau$}^2\Big)+78 \text{$\lambda_1$}^3\Big)-24 \text{$\lambda_t$}^2 \Big(3 \text{$\lambda_1$} \text{$\lambda_b$}^2+8 \text{$\lambda_3$} \text{$\lambda_4$}+8 \text{$\lambda_3$}^2+4 \Big(\text{$\lambda_4$}^2+\text{$\lambda_5$}^2\Big)-4 \text{$\lambda_b$}^4\Big)\Big)\Big].
\end{align}

\begin{align}
 16\pi^2\beta_{\lambda_4} &= 3 \text{$g_1$}^2 \Big(\text{$g_2$}^2-\text{$\lambda_4$}\Big)-9 \text{$g_2$}^2 \text{$\lambda_4$}+2 \text{$\lambda_4$} \Big(\text{$\lambda_1$}+\text{$\lambda_2$}+4 \text{$\lambda_3$}+2 \text{$\lambda_4$}+3 \text{$\lambda_b$}^2+\text{$\lambda_\tau$}^2\Big)\nonumber \\
  &+6 \text{$\lambda_t$}^2 \Big(\text{$\lambda_4$}+2 \text{$\lambda_b$}^2\Big)+8 \text{$\lambda_5$}^2 \nonumber \\
 &+ \frac{1}{384 \pi ^2}\Big[2 \text{$g_1$}^2 \Big(3 \text{$g_2$}^2 \Big(20 \text{$\lambda_1$}+20 \text{$\lambda_2$}+8 \text{$\lambda_3$}+51 \text{$\lambda_4$}+36 \text{$\lambda_b$}^2+44 \text{$\lambda_\tau$}^2+84 \text{$\lambda_t$}^2\Big)\nonumber \\
  &-168 \text{$g_2$}^4+\text{$\lambda_4$} \Big(48 \text{$\lambda_1$}+48 \text{$\lambda_2$}+48 \text{$\lambda_3$}+96 \text{$\lambda_4$}+25 \text{$\lambda_b$}^2+75 \text{$\lambda_\tau$}^2\Big)\nonumber \\
  &+\text{$\lambda_t$}^2 \Big(85 \text{$\lambda_4$}+16 \text{$\lambda_b$}^2\Big)+192 \text{$\lambda_5$}^2\Big)+\text{$g_1$}^4 \Big(471 \text{$\lambda_4$}-876 \text{$g_2$}^2\Big)\nonumber \\
  &+3 \Big(6 \text{$g_2$}^2 \Big(\text{$\lambda_4$} \Big(48 \text{$\lambda_3$}+24 \text{$\lambda_4$}+5 \Big(3 \text{$\lambda_b$}^2+\text{$\lambda_\tau$}^2+3 \text{$\lambda_t$}^2\Big)\Big)+72 \text{$\lambda_5$}^2\Big)-231 \text{$g_2$}^4 \text{$\lambda_4$}\nonumber \\
  &+4 \Big(-8 \text{$\lambda_b$}^2 \Big(-10 \text{$g_3$}^2 \text{$\lambda_4$}+3 \text{$\lambda_4$} (\text{$\lambda_1$}+2 \text{$\lambda_3$}+\text{$\lambda_4$})+6 \text{$\lambda_5$}^2\Big)\nonumber \\
  &-2 \text{$\lambda_t$}^2 \Big(3 \Big(4 \text{$\lambda_4$} (\text{$\lambda_2$}+2 \text{$\lambda_3$}+\text{$\lambda_4$})+\text{$\lambda_b$}^2 (8 \text{$\lambda_3$}+11 \text{$\lambda_4$})+8 \text{$\lambda_5$}^2+8 \text{$\lambda_b$}^4\Big)\nonumber \\
  &-8 \text{$g_3$}^2 \Big(5 \text{$\lambda_4$}+8 \text{$\lambda_b$}^2\Big)\Big)-4 \text{$\lambda_5$}^2 (12 (\text{$\lambda_1$}+\text{$\lambda_2$}+2 \text{$\lambda_3$})+13 \text{$\lambda_4$})\nonumber \\
  &-2 \text{$\lambda_4$} \Big(20 \text{$\lambda_1$} (2 \text{$\lambda_3$}+\text{$\lambda_4$})+7 \text{$\lambda_1$}^2+20 \text{$\lambda_2$} (2 \text{$\lambda_3$}+\text{$\lambda_4$})+7 \text{$\lambda_2$}^2+28 \text{$\lambda_3$} (\text{$\lambda_3$}+\text{$\lambda_4$})\Big)\nonumber \\
  &-8 \text{$\lambda_\tau$}^2 \Big(\text{$\lambda_4$} (\text{$\lambda_1$}+2 \text{$\lambda_3$}+\text{$\lambda_4$})+2 \text{$\lambda_5$}^2\Big)-3 \text{$\lambda_t$}^4 \Big(9 \text{$\lambda_4$}+16 \text{$\lambda_b$}^2\Big)-27 \text{$\lambda_4$} \text{$\lambda_b$}^4-9 \text{$\lambda_4$} \text{$\lambda_\tau$}^4\Big)\Big)\Big].
\end{align}

\begin{align}
 16 \pi^2\beta_{\lambda_5} & = \text{$\lambda_5$}  \Big(-3 \text{$g_1$}^2-9 \text{$g_2$}^2+2 \Big(\text{$\lambda_1$}+\text{$\lambda_2$}+4 \text{$\lambda_3$}+6 \text{$\lambda_4$}+3 \text{$\lambda_b$}^2+\text{$\lambda_\tau$}^2+3 \text{$\lambda_t$}^2\Big)\Big) \nonumber \\
  &+ \frac{1}{384 \pi ^2}\Big[\text{$\lambda_5$} \Big(2 \text{$g_1$}^2 \Big(57 \text{$g_2$}^2-24 \text{$\lambda_1$}-24 \text{$\lambda_2$}+192 \text{$\lambda_3$}+288 \text{$\lambda_4$}+25 \text{$\lambda_b$}^2+75 \text{$\lambda_\tau$}^2+85 \text{$\lambda_t$}^2\Big)\nonumber \\
  &+471 \text{$g_1$}^4+3 \Big(6 \text{$g_2$}^2 \Big(48 \text{$\lambda_3$}+96 \text{$\lambda_4$}+5 \Big(3 \text{$\lambda_b$}^2+\text{$\lambda_\tau$}^2+3 \text{$\lambda_t$}^2\Big)\Big)-231 \text{$g_2$}^4\nonumber \\
  &+4 \Big(-8 \text{$\lambda_b$}^2 \Big(-10 \text{$g_3$}^2+3 \text{$\lambda_1$}+6 \text{$\lambda_3$}+9 \text{$\lambda_4$}\Big)\nonumber \\
  &-2 \text{$\lambda_t$}^2 \Big(-40 \text{$g_3$}^2+12 (\text{$\lambda_2$}+2 \text{$\lambda_3$}+3 \text{$\lambda_4$})+33 \text{$\lambda_b$}^2\Big)-8 \text{$\lambda_4$} (11 (\text{$\lambda_1$}+\text{$\lambda_2$})+19 \text{$\lambda_3$})\nonumber \\
  &-8 \text{$\lambda_3$} (10 (\text{$\lambda_1$}+\text{$\lambda_2$})+7 \text{$\lambda_3$})-8 \text{$\lambda_\tau$}^2 (\text{$\lambda_1$}+2 \text{$\lambda_3$}+3 \text{$\lambda_4$})-64 \text{$\lambda_4$}^2-3 \text{$\lambda_b$}^4-\text{$\lambda_\tau$}^4-3 \text{$\lambda_t$}^4\Big)\nonumber \\
  &-56 \Big(\text{$\lambda_1$}^2+\text{$\lambda_2$}^2\Big)+48 \text{$\lambda_5$}^2\Big)\Big)\Big].
\end{align}

The $\beta$ functions for the Yukawa couplings $\lambda_t$, $\lambda_b$ and $\lambda_\tau$ are given by:

\begin{align}
 16 \pi^2 \beta_{\lambda_t} &= \frac{1}{12} \text{$\lambda_t$} \Big(-17 \text{$g_1$}(t)^2-27 \text{$g_2$}(t)^2+6 \Big(-16 \text{$g_3$}(t)^2+\text{$\lambda_b$}^2+9 \text{$\lambda_t$}^2\Big)\Big) \nonumber \\
 & -\frac{1}{6912 \pi ^2} \Big[ \text{$\lambda_t$} \Big(-2534 \text{$g_1$}^4+3 \text{$g_1$}^2 \Big(108 \text{$g_2$}^2-304 \text{$g_3$}^2+41 \text{$\lambda_b$}^2-1179 \text{$\lambda_t$}^2\Big) \nonumber \\
 & +9 \Big(252 \text{$g_2$}^4-9 \text{$g_2$}^2 \Big(48 \text{$g_3$}^2+11 \text{$\lambda_b$}^2+75 \text{$\lambda_t$}^2\Big)+4 \Big(1296 \text{$g_3$}^4-16 \text{$g_3$}^2 \Big(4 \text{$\lambda_b$}^2+27 \text{$\lambda_t$}^2\Big) \nonumber \\
 & -3 \Big(6 \text{$\lambda_2$}^2-24 \text{$\lambda_2$} \text{$\lambda_t$}^2+4 \text{$\lambda_3$}^2+4 \text{$\lambda_3$} \text{$\lambda_4$}-8 \text{$\lambda_3$} \text{$\lambda_b$}^2+4 \text{$\lambda_4$}^2+8 \text{$\lambda_4$} \text{$\lambda_b$}^2 \nonumber \\
 & +6 \text{$\lambda_5$}^2-10 \text{$\lambda_b$}^4-3 \text{$\lambda_b$}^2 \text{$\lambda_\tau$}^2-10 \text{$\lambda_b$}^2 \text{$\lambda_t$}^2-48 \text{$\lambda_t$}^4\Big)\Big)\Big)\Big)\Big]
\end{align}

\begin{align}
 16\pi^2\beta_{\lambda_b} &= \frac{1}{12} \text{$\lambda_b$} \Big(-5 \text{$g_1$}^2-27 \text{$g_2$}^2+6 \Big(-16 \text{$g_3$}^2+9 \text{$\lambda_b$}^2+2 \text{$\lambda_\tau$}^2+\text{$\lambda_t$}^2\Big)\Big)\nonumber \\
 & -\frac{1}{6912 \pi ^2}\Big[\text{$\lambda_b$} \Big(226 \text{$g_1$}^4+3 \text{$g_1$}^2 \Big(324 \text{$g_2$}^2-496 \text{$g_3$}^2-711 \text{$\lambda_b$}^2-450 \text{$\lambda_\tau$}^2+53 \text{$\lambda_t$}^2\Big)\nonumber \\ 
 & +9 \Big(252 \text{$g_2$}^4-9 \text{$g_2$}^2 \Big(48 \text{$g_3$}^2+75 \text{$\lambda_b$}^2+10 \text{$\lambda_\tau$}^2+11 \text{$\lambda_t$}^2\Big) \nonumber \\
 & +4 \Big(1296 \text{$g_3$}^4-16 \text{$g_3$}^2 \Big(27 \text{$\lambda_b$}^2+4 \text{$\lambda_t$}^2\Big)-3 \Big(6 \text{$\lambda_1$}^2-24 \text{$\lambda_1$} \text{$\lambda_b$}^2+4 \text{$\lambda_3$}^2+4 \text{$\lambda_3$} \text{$\lambda_4$} \nonumber \\
 & -8 \text{$\lambda_3$} \text{$\lambda_t$}^2+4 \text{$\lambda_4$}^2+8 \text{$\lambda_4$} \text{$\lambda_t$}^2+6 \text{$\lambda_5$}^2-48 \text{$\lambda_b$}^4-9 \text{$\lambda_b$}^2 \text{$\lambda_\tau$}^2-10 \text{$\lambda_b$}^2 \text{$\lambda_t$}^2-9 \text{$\lambda_\tau$}^4-10 \text{$\lambda_t$}^4\Big)\Big)\Big)\Big)\Big]
\end{align}

\begin{align}
 16\pi^2\beta_{\lambda_\tau} &=  \frac{1}{4} \text{$\lambda_\tau$} \Big(-15 \text{$g_1$}^2-9 \text{$g_2$}^2+12 \text{$\lambda_b$}^2+10 \text{$\lambda_\tau$}^2\Big) \nonumber \\
 & -\frac{1}{768 \pi ^2}\Big[\text{$\lambda_\tau$} \Big(-966 \text{$g_1$}^4-\text{$g_1$}^2 \Big(108 \text{$g_2$}^2+50 \text{$\lambda_b$}^2+537 \text{$\lambda_\tau$}^2\Big) \nonumber \\
  & +3 \Big(84 \text{$g_2$}^4-15 \text{$g_2$}^2 \Big(6 \text{$\lambda_b$}^2+11 \text{$\lambda_\tau$}^2\Big)-4 \Big(80 \text{$g_3$}^2 \text{$\lambda_b$}^2+6 \text{$\lambda_1$}^2-24 \text{$\lambda_1$} \text{$\lambda_\tau$}^2 \nonumber \\
  & +4 \text{$\lambda_3$}^2+4 \text{$\lambda_3$} \text{$\lambda_4$}+4 \text{$\lambda_4$}^2+6 \text{$\lambda_5$}^2-27 \text{$\lambda_b$}^4-27 \text{$\lambda_b$}^2 \text{$\lambda_\tau$}^2-9 \text{$\lambda_b$}^2 \text{$\lambda_t$}^2-12 \text{$\lambda_\tau$}^4\Big)\Big)\Big)\Big]
\end{align}

\bibliographystyle{JHEP}
\bibliography{bibi}

\providecommand{\href}[2]{#2}\begingroup\raggedright\begin{thebibliography}{10}

\bibitem{Aad:2012tfa}
{\scshape ATLAS} collaboration, G.~Aad et~al., \emph{{Observation of a new
  particle in the search for the Standard Model Higgs boson with the ATLAS
  detector at the LHC}},
  \href{https://doi.org/10.1016/j.physletb.2012.08.020}{\emph{Phys. Lett.}
  {\bfseries B716} (2012) 1} [\href{https://arxiv.org/abs/1207.7214}{{\ttfamily
  1207.7214}}].

\bibitem{Chatrchyan:2012xdj}
{\scshape CMS} collaboration, S.~Chatrchyan et~al., \emph{{Observation of a new
  boson at a mass of 125 GeV with the CMS experiment at the LHC}},
  \href{https://doi.org/10.1016/j.physletb.2012.08.021}{\emph{Phys. Lett.}
  {\bfseries B716} (2012) 30}
  [\href{https://arxiv.org/abs/1207.7235}{{\ttfamily 1207.7235}}].

\bibitem{EliasMiro:2011aa}
J.~Elias-Miro, J.~R. Espinosa, G.~F. Giudice, G.~Isidori, A.~Riotto and
  A.~Strumia, \emph{{Higgs mass implications on the stability of the
  electroweak vacuum}},
  \href{https://doi.org/10.1016/j.physletb.2012.02.013}{\emph{Phys. Lett.}
  {\bfseries B709} (2012) 222}
  [\href{https://arxiv.org/abs/1112.3022}{{\ttfamily 1112.3022}}].

\bibitem{Branchina:2013jra}
V.~Branchina and E.~Messina, \emph{{Stability, Higgs Boson Mass and New
  Physics}}, \href{https://doi.org/10.1103/PhysRevLett.111.241801}{\emph{Phys.
  Rev. Lett.} {\bfseries 111} (2013) 241801}
  [\href{https://arxiv.org/abs/1307.5193}{{\ttfamily 1307.5193}}].

\bibitem{Buttazzo:2013uya}
D.~Buttazzo, G.~Degrassi, P.~P. Giardino, G.~F. Giudice, F.~Sala, A.~Salvio
  et~al., \emph{{Investigating the near-criticality of the Higgs boson}},
  \href{https://doi.org/10.1007/JHEP12(2013)089}{\emph{JHEP} {\bfseries 12}
  (2013) 089} [\href{https://arxiv.org/abs/1307.3536}{{\ttfamily 1307.3536}}].

\bibitem{Shaposhnikov:2009pv}
M.~Shaposhnikov and C.~Wetterich, \emph{{Asymptotic safety of gravity and the
  Higgs boson mass}},
  \href{https://doi.org/10.1016/j.physletb.2009.12.022}{\emph{Phys.Lett.}
  {\bfseries B683} (2010) 196}
  [\href{https://arxiv.org/abs/0912.0208}{{\ttfamily 0912.0208}}].

\bibitem{Holthausen:2011aa}
M.~Holthausen, K.~S. Lim and M.~Lindner, \emph{{Planck scale Boundary
  Conditions and the Higgs Mass}},
  \href{https://doi.org/10.1007/JHEP02(2012)037}{\emph{JHEP} {\bfseries 1202}
  (2012) 037} [\href{https://arxiv.org/abs/1112.2415}{{\ttfamily 1112.2415}}].

\bibitem{Weinberg:1980gg}
S.~Weinberg, \emph{{Ultraviolet divergencies in theories of quantum
  gravitation}},  in \emph{{General Relativity: An Einstein centenary survey}}
  (S.~Hawking and W.~Israel, eds.), pp.~790--831.
\newblock Cambride University Press, 1979.

\bibitem{Wetterich:1992yh}
C.~Wetterich, \emph{{Exact evolution equation for the effective potential}},
  \href{https://doi.org/10.1016/0370-2693(93)90726-X}{\emph{Phys.Lett.}
  {\bfseries B301} (1993) 90}.

\bibitem{Percacci:2011fr}
R.~Percacci, \emph{{A Short introduction to asymptotic safety}},
  \href{https://arxiv.org/abs/1110.6389}{{\ttfamily 1110.6389}}.

\bibitem{Bond:2016dvk}
A.~D. Bond and D.~F. Litim, \emph{{Theorems for Asymptotic Safety of Gauge
  Theories}},  \href{https://arxiv.org/abs/1608.00519}{{\ttfamily 1608.00519}}.

\bibitem{Bond:2017wut}
A.~D. Bond, G.~Hiller, K.~Kowalska and D.~F. Litim, \emph{{Directions for model
  building from asymptotic safety}},
  \href{https://doi.org/10.1007/JHEP08(2017)004}{\emph{JHEP} {\bfseries 08}
  (2017) 004} [\href{https://arxiv.org/abs/1702.01727}{{\ttfamily
  1702.01727}}].

\bibitem{Pelaggi:2017abg}
G.~M. Pelaggi, A.~D. Plascencia, A.~Salvio, F.~Sannino, J.~Smirnov and
  A.~Strumia, \emph{{Asymptotically Safe Standard Model Extensions?}},
  \href{https://arxiv.org/abs/1708.00437}{{\ttfamily 1708.00437}}.

\bibitem{Bond:2017lnq}
A.~D. Bond and D.~F. Litim, \emph{{More asymptotic safety guaranteed}},
  \href{https://arxiv.org/abs/1707.04217}{{\ttfamily 1707.04217}}.

\bibitem{Bond:2018oco}
A.~D. Bond and D.~F. Litim, \emph{{Price of Asymptotic Safety}},
  \href{https://arxiv.org/abs/1801.08527}{{\ttfamily 1801.08527}}.

\bibitem{Barducci:2018ysr}
D.~Barducci, M.~Fabbrichesi, C.~M. Nieto, R.~Percacci and V.~Skrinjar,
  \emph{{In search of a UV completion of the Standard Model - 378.000 models
  that don't work}},  \href{https://arxiv.org/abs/1807.05584}{{\ttfamily
  1807.05584}}.

\bibitem{Mann:2017wzh}
R.~Mann, J.~Meffe, F.~Sannino, T.~Steele, Z.-W. Wang and C.~Zhang,
  \emph{{Asymptotically Safe Standard Model via Vectorlike Fermions}},
  \href{https://doi.org/10.1103/PhysRevLett.119.261802}{\emph{Phys. Rev. Lett.}
  {\bfseries 119} (2017) 261802}
  [\href{https://arxiv.org/abs/1707.02942}{{\ttfamily 1707.02942}}].

\bibitem{Bednyakov:2015sca}
A.~V. Bednyakov, B.~A. Kniehl, A.~F. Pikelner and O.~L. Veretin,
  \emph{{Stability of the Electroweak Vacuum: Gauge Independence and Advanced
  Precision}},
  \href{https://doi.org/10.1103/PhysRevLett.115.201802}{\emph{Phys. Rev. Lett.}
  {\bfseries 115} (2015) 201802}
  [\href{https://arxiv.org/abs/1507.08833}{{\ttfamily 1507.08833}}].

\bibitem{Lee:1973iz}
T.~D. Lee, \emph{{A Theory of Spontaneous T Violation}},
  \href{https://doi.org/10.1103/PhysRevD.8.1226}{\emph{Phys. Rev.} {\bfseries
  D8} (1973) 1226}.

\bibitem{Branco2012}
G.~C. Branco, P.~M. Ferreira, L.~Lavoura, M.~N. Rebelo, M.~Sher and J.~P.
  Silva, \emph{{Theory and phenomenology of two-Higgs-doublet models}},
  \href{https://doi.org/10.1016/j.physrep.2012.02.002}{\emph{Phys. Rept.}
  {\bfseries 516} (2012) 1} [\href{https://arxiv.org/abs/1106.0034}{{\ttfamily
  1106.0034}}].

\bibitem{Paschos:1976ay}
E.~A. Paschos, \emph{{Diagonal Neutral Currents}},
  \href{https://doi.org/10.1103/PhysRevD.15.1966}{\emph{Phys. Rev.} {\bfseries
  D15} (1977) 1966}.

\bibitem{Glashow:1976nt}
S.~L. Glashow and S.~Weinberg, \emph{{Natural Conservation Laws for Neutral
  Currents}}, \href{https://doi.org/10.1103/PhysRevD.15.1958}{\emph{Phys. Rev.}
  {\bfseries D15} (1977) 1958}.

\bibitem{Aoki:2009ha}
M.~Aoki, S.~Kanemura, K.~Tsumura and K.~Yagyu, \emph{{Models of Yukawa
  interaction in the two Higgs doublet model, and their collider
  phenomenology}},
  \href{https://doi.org/10.1103/PhysRevD.80.015017}{\emph{Phys. Rev.}
  {\bfseries D80} (2009) 015017}
  [\href{https://arxiv.org/abs/0902.4665}{{\ttfamily 0902.4665}}].

\bibitem{Deshpande:1977rw}
N.~G. Deshpande and E.~Ma, \emph{{Pattern of Symmetry Breaking with Two Higgs
  Doublets}}, \href{https://doi.org/10.1103/PhysRevD.18.2574}{\emph{Phys. Rev.}
  {\bfseries D18} (1978) 2574}.

\bibitem{Branchina:2018qlf}
V.~Branchina, F.~Contino and P.~M. Ferreira, \emph{{Electroweak vacuum lifetime
  in two Higgs doublet models}},
  \href{https://arxiv.org/abs/1807.10802}{{\ttfamily 1807.10802}}.

\bibitem{Ferreira:2004yd}
P.~M. Ferreira, R.~Santos and A.~Barroso, \emph{{Stability of the tree-level
  vacuum in two Higgs doublet models against charge or CP spontaneous
  violation}}, \href{https://doi.org/10.1016/j.physletb.2004.10.022,
  10.1016/j.physletb.2005.09.074}{\emph{Phys. Lett.} {\bfseries B603} (2004)
  219} [\href{https://arxiv.org/abs/hep-ph/0406231}{{\ttfamily
  hep-ph/0406231}}].

\bibitem{Barroso:2005sm}
A.~Barroso, P.~M. Ferreira and R.~Santos, \emph{{Charge and CP symmetry
  breaking in two Higgs doublet models}},
  \href{https://doi.org/10.1016/j.physletb.2005.11.031}{\emph{Phys. Lett.}
  {\bfseries B632} (2006) 684}
  [\href{https://arxiv.org/abs/hep-ph/0507224}{{\ttfamily hep-ph/0507224}}].

\bibitem{Barroso:2007rr}
A.~Barroso, P.~M. Ferreira and R.~Santos, \emph{{Neutral minima in two-Higgs
  doublet models}},
  \href{https://doi.org/10.1016/j.physletb.2007.07.010}{\emph{Phys. Lett.}
  {\bfseries B652} (2007) 181}
  [\href{https://arxiv.org/abs/hep-ph/0702098}{{\ttfamily hep-ph/0702098}}].

\bibitem{Barroso:2013zxa}
A.~Barroso, P.~M. Ferreira, R.~Santos, M.~Sher and J.~P. Silva, \emph{{2HDM at
  the LHC - the story so far}},  in \emph{{Proceedings, 1st Toyama
  International Workshop on Higgs as a Probe of New Physics 2013 (HPNP2013):
  Toyama, Japan, February 13-16, 2013}}, 2013,
  \href{https://arxiv.org/abs/1304.5225}{{\ttfamily 1304.5225}},
  \href{http://inspirehep.net/record/1228915/files/arXiv:1304.5225.pdf}{http://inspirehep.net/record/1228915/files/arXiv:1304.5225.pdf}.

\bibitem{Coleppa:2013dya}
B.~Coleppa, F.~Kling and S.~Su, \emph{{Constraining Type II 2HDM in Light of
  LHC Higgs Searches}},
  \href{https://doi.org/10.1007/JHEP01(2014)161}{\emph{JHEP} {\bfseries 01}
  (2014) 161} [\href{https://arxiv.org/abs/1305.0002}{{\ttfamily 1305.0002}}].

\bibitem{Sirunyan:2018koj}
{\scshape CMS} collaboration, A.~M. Sirunyan et~al., \emph{{Combined
  measurements of Higgs boson couplings in proton-proton collisions at
  $\sqrt{s}=$ 13 TeV}},  \href{https://arxiv.org/abs/1809.10733}{{\ttfamily
  1809.10733}}.

\bibitem{Eberhardt:2017ulj}
O.~Eberhardt, \emph{{Two-Higgs-doublet model fits with HEPfit}},  in
  \emph{{2017 European Physical Society Conference on High Energy Physics
  (EPS-HEP 2017) Venice, Italy, July 5-12, 2017}}, 2017,
  \href{https://arxiv.org/abs/1709.09414}{{\ttfamily 1709.09414}},
  \href{http://inspirehep.net/record/1626106/files/arXiv:1709.09414.pdf}{http://inspirehep.net/record/1626106/files/arXiv:1709.09414.pdf}.

\bibitem{Chowdhury:2017aav}
D.~Chowdhury and O.~Eberhardt, \emph{{Update of Global Two-Higgs-Doublet Model
  Fits}},  \href{https://arxiv.org/abs/1711.02095}{{\ttfamily 1711.02095}}.

\bibitem{Dorsch:2016tab}
G.~C. Dorsch, S.~J. Huber, K.~Mimasu and J.~M. No, \emph{{Hierarchical versus
  degenerate 2HDM: The LHC run 1 legacy at the onset of run 2}},
  \href{https://doi.org/10.1103/PhysRevD.93.115033}{\emph{Phys. Rev.}
  {\bfseries D93} (2016) 115033}
  [\href{https://arxiv.org/abs/1601.04545}{{\ttfamily 1601.04545}}].

\bibitem{Cheon:2012rh}
H.~S. Cheon and S.~K. Kang, \emph{{Constraining parameter space in type-II
  two-Higgs doublet model in light of a 126 GeV Higgs boson}},
  \href{https://doi.org/10.1007/JHEP09(2013)085}{\emph{JHEP} {\bfseries 09}
  (2013) 085} [\href{https://arxiv.org/abs/1207.1083}{{\ttfamily 1207.1083}}].

\bibitem{Eberhardt:2013uba}
O.~Eberhardt, U.~Nierste and M.~Wiebusch, \emph{{Status of the
  two-Higgs-doublet model of type II}},
  \href{https://doi.org/10.1007/JHEP07(2013)118}{\emph{JHEP} {\bfseries 07}
  (2013) 118} [\href{https://arxiv.org/abs/1305.1649}{{\ttfamily 1305.1649}}].

\bibitem{Chowdhury:2015yja}
D.~Chowdhury and O.~Eberhardt, \emph{{Global fits of the two-loop renormalized
  Two-Higgs-Doublet model with soft Z$_{2}$ breaking}},
  \href{https://doi.org/10.1007/JHEP11(2015)052}{\emph{JHEP} {\bfseries 11}
  (2015) 052} [\href{https://arxiv.org/abs/1503.08216}{{\ttfamily
  1503.08216}}].

\bibitem{Chakrabarty:2014aya}
N.~Chakrabarty, U.~K. Dey and B.~Mukhopadhyaya, \emph{{High-scale validity of a
  two-Higgs doublet scenario: a study including LHC data}},
  \href{https://doi.org/10.1007/JHEP12(2014)166}{\emph{JHEP} {\bfseries 12}
  (2014) 166} [\href{https://arxiv.org/abs/1407.2145}{{\ttfamily 1407.2145}}].

\bibitem{Lee:1977yc}
B.~W. Lee, C.~Quigg and H.~B. Thacker, \emph{{The Strength of Weak Interactions
  at Very High-Energies and the Higgs Boson Mass}},
  \href{https://doi.org/10.1103/PhysRevLett.38.883}{\emph{Phys. Rev. Lett.}
  {\bfseries 38} (1977) 883}.

\bibitem{Grinstein:2015rtl}
B.~Grinstein, C.~W. Murphy and P.~Uttayarat, \emph{{One-loop corrections to the
  perturbative unitarity bounds in the CP-conserving two-Higgs doublet model
  with a softly broken $ {\mathrm{\mathbb{Z}}}_2 $ symmetry}},
  \href{https://doi.org/10.1007/JHEP06(2016)070}{\emph{JHEP} {\bfseries 06}
  (2016) 070} [\href{https://arxiv.org/abs/1512.04567}{{\ttfamily
  1512.04567}}].

\bibitem{ATLAS:2014yka}
{\scshape ATLAS} collaboration, T.~A. collaboration, \emph{{Updated coupling
  measurements of the Higgs boson with the ATLAS detector using up to 25
  fb$^{-1}$ of proton-proton collision data}}, .

\bibitem{CMS:2014ega}
{\scshape CMS} collaboration, C.~Collaboration, \emph{{Precise determination of
  the mass of the Higgs boson and studies of the compatibility of its couplings
  with the standard model}}, .

\bibitem{ATLAS:2018uoi}
{\scshape ATLAS} collaboration, T.~A. collaboration, \emph{{Combined
  measurement of differential and inclusive total cross sections in the $H
  \rightarrow \gamma \gamma$ and the $H \rightarrow ZZ^* \rightarrow 4\ell$
  decay channels at $\sqrt{s} = 13$ TeV with the ATLAS detector}}, .

\bibitem{CMS:2018lkl}
{\scshape CMS} collaboration, C.~Collaboration, \emph{{Combined measurements of
  the Higgs boson's couplings at $\sqrt{s}=13$ TeV}}, .

\bibitem{Enomoto:2015wbn}
T.~Enomoto and R.~Watanabe, \emph{{Flavor constraints on the Two Higgs Doublet
  Models of $Z_2$ symmetric and aligned types}},
  \href{https://arxiv.org/abs/1511.05066}{{\ttfamily 1511.05066}}.

\bibitem{Misiak:2017bgg}
M.~Misiak and M.~Steinhauser, \emph{{Weak radiative decays of the B meson and
  bounds on $M_{H^\pm }$ in the Two-Higgs-Doublet Model}},
  \href{https://doi.org/10.1140/epjc/s10052-017-4776-y}{\emph{Eur. Phys. J.}
  {\bfseries C77} (2017) 201}
  [\href{https://arxiv.org/abs/1702.04571}{{\ttfamily 1702.04571}}].

\bibitem{Haller:2018nnx}
J.~Haller, A.~Hoecker, R.~Kogler, K.~Mönig, T.~Peiffer and J.~Stelzer,
  \emph{{Update of the global electroweak fit and constraints on
  two-Higgs-doublet models}},
  \href{https://doi.org/10.1140/epjc/s10052-018-6131-3}{\emph{Eur. Phys. J.}
  {\bfseries C78} (2018) 675}
  [\href{https://arxiv.org/abs/1803.01853}{{\ttfamily 1803.01853}}].

\bibitem{Gunion:2002zf}
J.~F. Gunion and H.~E. Haber, \emph{{The CP conserving two Higgs doublet model:
  The Approach to the decoupling limit}},
  \href{https://doi.org/10.1103/PhysRevD.67.075019}{\emph{Phys. Rev.}
  {\bfseries D67} (2003) 075019}
  [\href{https://arxiv.org/abs/hep-ph/0207010}{{\ttfamily hep-ph/0207010}}].

\bibitem{Das:2015mwa}
D.~Das and I.~Saha, \emph{{Search for a stable alignment limit in
  two-Higgs-doublet models}},
  \href{https://doi.org/10.1103/PhysRevD.91.095024}{\emph{Phys. Rev.}
  {\bfseries D91} (2015) 095024}
  [\href{https://arxiv.org/abs/1503.02135}{{\ttfamily 1503.02135}}].

\bibitem{Carena:2013ooa}
M.~Carena, I.~Low, N.~R. Shah and C.~E.~M. Wagner, \emph{{Impersonating the
  Standard Model Higgs Boson: Alignment without Decoupling}},
  \href{https://doi.org/10.1007/JHEP04(2014)015}{\emph{JHEP} {\bfseries 04}
  (2014) 015} [\href{https://arxiv.org/abs/1310.2248}{{\ttfamily 1310.2248}}].

\bibitem{Dev:2014yca}
P.~S. Bhupal~Dev and A.~Pilaftsis, \emph{{Maximally Symmetric Two Higgs Doublet
  Model with Natural Standard Model Alignment}},
  \href{https://doi.org/10.1007/JHEP11(2015)147,
  10.1007/JHEP12(2014)024}{\emph{JHEP} {\bfseries 12} (2014) 024}
  [\href{https://arxiv.org/abs/1408.3405}{{\ttfamily 1408.3405}}].

\bibitem{Staub:2008uz}
F.~Staub, \emph{{SARAH}},  \href{https://arxiv.org/abs/0806.0538}{{\ttfamily
  0806.0538}}.

\bibitem{Staub:2013tta}
F.~Staub, \emph{{SARAH 4 : A tool for (not only SUSY) model builders}},
  \href{https://doi.org/10.1016/j.cpc.2014.02.018}{\emph{Comput. Phys. Commun.}
  {\bfseries 185} (2014) 1773}
  [\href{https://arxiv.org/abs/1309.7223}{{\ttfamily 1309.7223}}].

\bibitem{Mohr:2015ccw}
P.~J. Mohr, D.~B. Newell and B.~N. Taylor, \emph{{CODATA Recommended Values of
  the Fundamental Physical Constants: 2014}},
  \href{https://doi.org/10.1103/RevModPhys.88.035009}{\emph{Rev. Mod. Phys.}
  {\bfseries 88} (2016) 035009}
  [\href{https://arxiv.org/abs/1507.07956}{{\ttfamily 1507.07956}}].

\bibitem{Agashe:2014kda}
{\scshape Particle Data Group} collaboration, K.~A. Olive et~al., \emph{{Review
  of Particle Physics}},
  \href{https://doi.org/10.1088/1674-1137/38/9/090001}{\emph{Chin. Phys.}
  {\bfseries C38} (2014) 090001}.

\bibitem{Falkowski:2015iwa}
A.~Falkowski, C.~Gross and O.~Lebedev, \emph{{A second Higgs from the Higgs
  portal}}, \href{https://doi.org/10.1007/JHEP05(2015)057}{\emph{JHEP}
  {\bfseries 05} (2015) 057}
  [\href{https://arxiv.org/abs/1502.01361}{{\ttfamily 1502.01361}}].

\bibitem{Ferreira:2012my}
P.~M. Ferreira, R.~Santos, M.~Sher and J.~P. Silva, \emph{{Could the LHC
  two-photon signal correspond to the heavier scalar in two-Higgs-doublet
  models?}}, \href{https://doi.org/10.1103/PhysRevD.85.035020}{\emph{Phys.
  Rev.} {\bfseries D85} (2012) 035020}
  [\href{https://arxiv.org/abs/1201.0019}{{\ttfamily 1201.0019}}].

\bibitem{Cheng:2015yfu}
X.-D. Cheng, Y.-D. Yang and X.-B. Yuan, \emph{{Revisiting $B_s \to \mu^+\mu^-$
  in the two-Higgs doublet models with $Z_2$ symmetry}},
  \href{https://doi.org/10.1140/epjc/s10052-016-3930-2}{\emph{Eur. Phys. J.}
  {\bfseries C76} (2016) 151}
  [\href{https://arxiv.org/abs/1511.01829}{{\ttfamily 1511.01829}}].

\bibitem{McDowall:2018ulq}
J.~McDowall and D.~J. Miller, \emph{{High Scale Boundary Conditions in Models
  with Two Higgs Doublets}},
  \href{https://arxiv.org/abs/1810.04518}{{\ttfamily 1810.04518}}.

\bibitem{Machacek:1983tz}
M.~E. Machacek and M.~T. Vaughn, \emph{{Two Loop Renormalization Group
  Equations in a General Quantum Field Theory. 1. Wave Function
  Renormalization}},
  \href{https://doi.org/10.1016/0550-3213(83)90610-7}{\emph{Nucl.Phys.}
  {\bfseries B222} (1983) 83}.

\bibitem{Machacek:1983fi}
M.~E. Machacek and M.~T. Vaughn, \emph{{Two Loop Renormalization Group
  Equations in a General Quantum Field Theory. 2. Yukawa Couplings}},
  \href{https://doi.org/10.1016/0550-3213(84)90533-9}{\emph{Nucl.Phys.}
  {\bfseries B236} (1984) 221}.

\bibitem{Machacek:1984zw}
M.~E. Machacek and M.~T. Vaughn, \emph{{Two Loop Renormalization Group
  Equations in a General Quantum Field Theory. 3. Scalar Quartic Couplings}},
  \href{https://doi.org/10.1016/0550-3213(85)90040-9}{\emph{Nucl.Phys.}
  {\bfseries B249} (1985) 70}.

\end{thebibliography}\endgroup
\end{document}